\documentclass[%
aip,
floatfix,
rsi,
 amsmath,amssymb,
reprint,%
]{revtex4-1}
\usepackage{graphicx}
\usepackage{dcolumn}
\usepackage{bm}
\usepackage[utf8]{inputenc}
\usepackage[T1]{fontenc}
\usepackage{color}
\usepackage{textcomp}
\usepackage{amsmath}
\usepackage{booktabs}

\begin{document}
\preprint{Si5 manuscript v8}

\title{A simplified cryogenic optical resonator apparatus providing ultra-low frequency drift} 

\author{Eugen Wiens}
 \affiliation{Institut für Experimentalphysik, Heinrich-Heine-Universität
Düsseldorf, 40225 Düsseldorf, Germany}

\author{Chang Jian Kwong}%
\affiliation{Institut für Experimentalphysik, Heinrich-Heine-Universität
Düsseldorf, 40225 Düsseldorf, Germany
}%

\author{Timo Müller}%
\affiliation{Siltronic AG, Johannes-Hess-Straße 24, 84489 Burghausen, Germany
}%

\author{Stephan Schiller}%
\email{step.schiller@hhu.de.}
\affiliation{Institut für Experimentalphysik, Heinrich-Heine-Universität
Düsseldorf, 40225 Düsseldorf, Germany
}%

\date{\today}

\begin{abstract}
A system providing an optical frequency with an instability comparable to that of a hydrogen maser is presented. It consists of a $5$~cm long, vertically oriented silicon optical resonator operated at temperatures between $1.5$~K and 3.6~K in a closed-cycle cryostat with low-temperature Joule-Thomson stage. We show that with a standard cryostat, a simple cryogenic optomechanical setup, no active or passive vibration isolation, a minimum frequency instability of $2.5\times10^{-15}$ at $\tau=1500$~s integration time can be reached. The influence of pulse-tube vibrations was minimized by using a resonator designed for low acceleration sensitivity. With reduced optical laser power and interrogation duty cycle an ultra-low fractional frequency drift of $-2.6\times10^{-19}$/s is reached. At $3.5$~K the resonator frequency exhibits a vanishing thermal sensitivity and an ultra-small temperature derivative $8.5\times10^{-12}/\mathrm{K}^{2}$. These are favorable properties that should lead to high performance also in simpler cryostats not equipped with a Joule-Thomson stage. 
\end{abstract}


\maketitle 

\section{Introduction}

Optical resonators play an important role in the generation of laser
light with ultra-stable frequency. They are essential to the field
of optical atomic clocks, where they are utilized for the pre-stabilization
of the laser wave used for the interrogation of ultra-narrow atomic
transitions \cite{Ludlow2015-1,Derevianko2011,Poli2013}. Other applications
are in gravitational wave detectors \cite{Will2014,Adhikari2014}
and for tests of fundamental physics \cite{Braxmaier2002,Wiens2016,Chen2016,Tobar2010,Eisele2009,Nagel2015}.

The most common type of resonator consists of a hollow spacer of length
$L$ which introduces a fixed separation between two mirrors that
are optically contacted to its end surfaces. The design of the spacer
geometry and support is usually optimized to reduce length variations
produced by environmental vibrations. To counteract thermally induced
variations in length, the resonators are usually made of materials
that exhibit a particularly low thermal expansion coefficient at the
desired operational temperature. Ultra-low-expansion glass (ULE) is
today the most common material for use at or near room temperature.
Another promising material is the ceramic Nexcera \cite{Takahashi2012,Hosaka2013,Ito2017}.
However, the Brownian motion imposes a  fundamental limitation to their length
stability \cite{Numata2004a}. This is on the order of $1\times10^{-15}$ for room-temperature resonators with a typical length of $ \le $10~cm \cite{Ludlow2007,Webster2008, Chen2014, Swierad2016,Davila2017}.
The instability was successfully lowered to $8\times10^{-17}$ using a $48$~cm long resonator\cite{Haefner2015}.   
Furthermore, room-temperature resonators made of ULE material suffer from drift. The drift rates vary substantially
between units, with one of the smallest values being $1.6\times10^{-17}$/s (Ref.~\onlinecite{Ito2017}). One approach for reducing both limitations is the
operation of the resonators at cryogenic temperatures \cite{Matei2017,Zhang2017,Robinson2019,Seel1997,Kessler2012a,Hagemann2014,Wiens2014,Wiens2016,Braxmaier2002,Muller2002}.

Here, we present a cryogenic single-crystal silicon resonator developed
for low vibration sensitivity and frequency stability comparable to that of a hydrogen maser, operated
in a cryogenic system of moderate complexity. In order to reduce the
resonator manufacturing cost, we simplified the design to a cylindrical
shape. A fairly complete characterization of the resonator was possible
using a system composed of a stable interrogation laser, a frequency
comb and a hydrogen maser.
\begin{figure*}[htb]
	\begin{centering}
	 \includegraphics[width=0.48\textwidth]{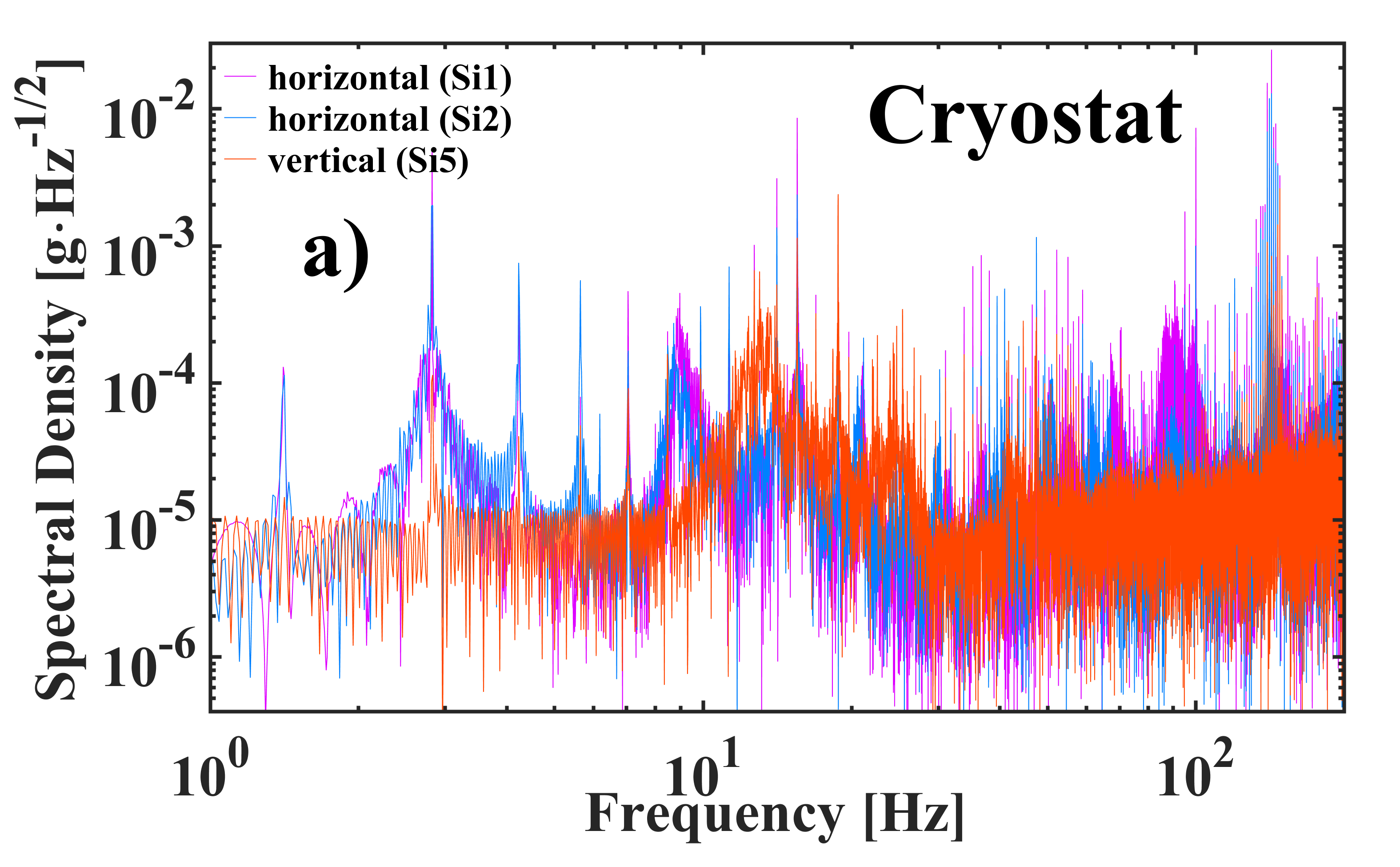}
	 \includegraphics[width=0.48\textwidth]{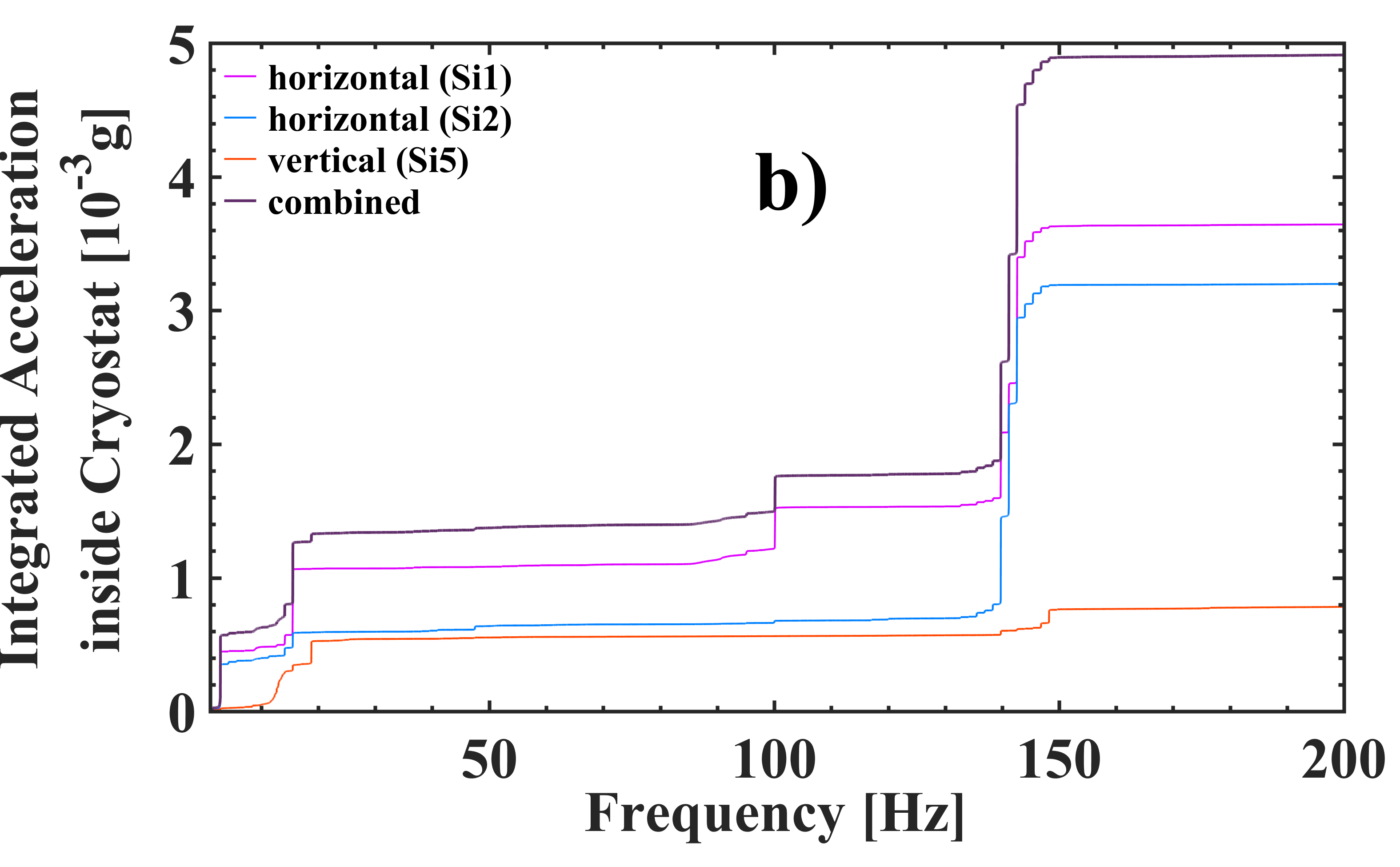}
	 \includegraphics[width=0.48\textwidth]{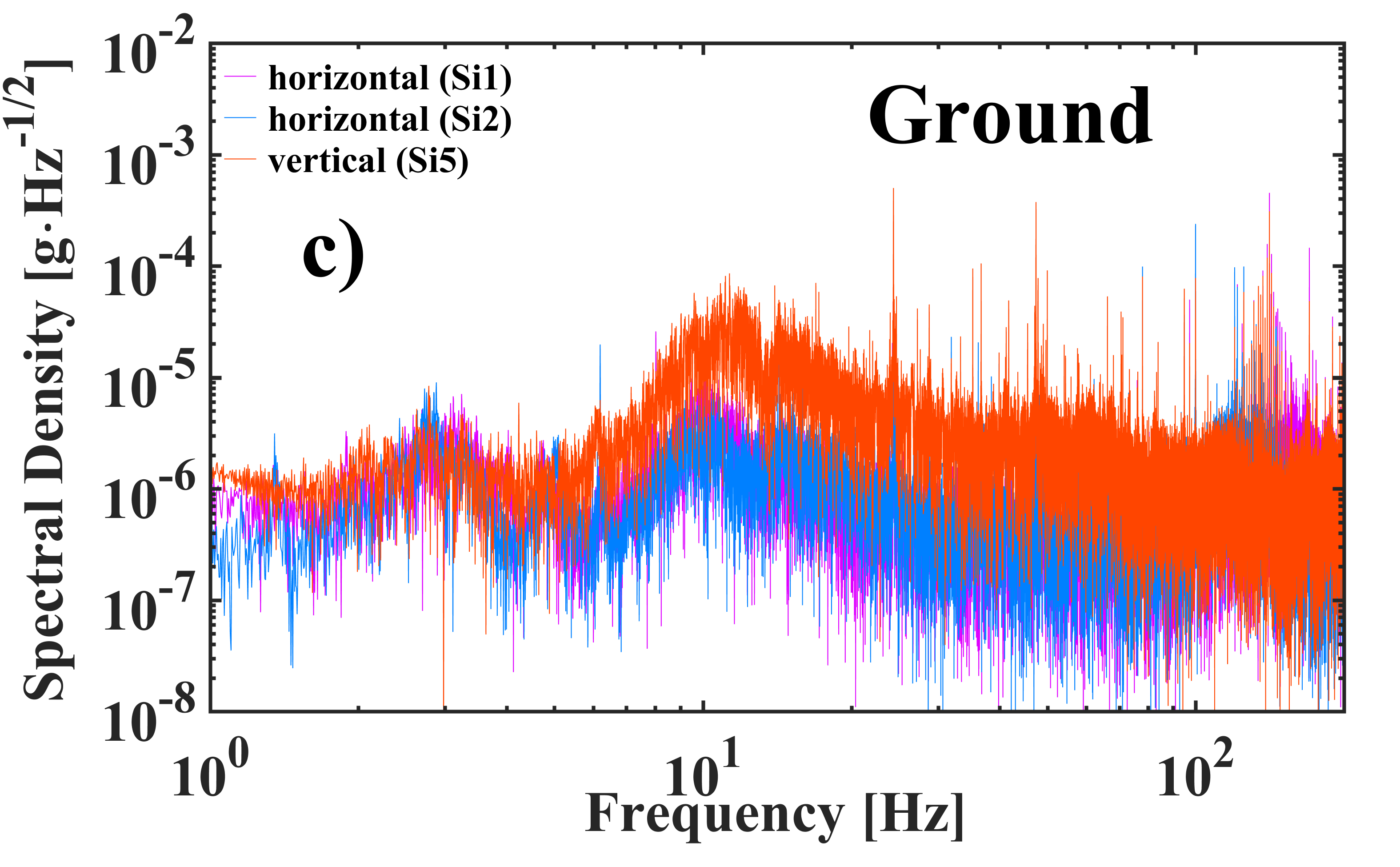}
	 \includegraphics[width=0.48\textwidth]{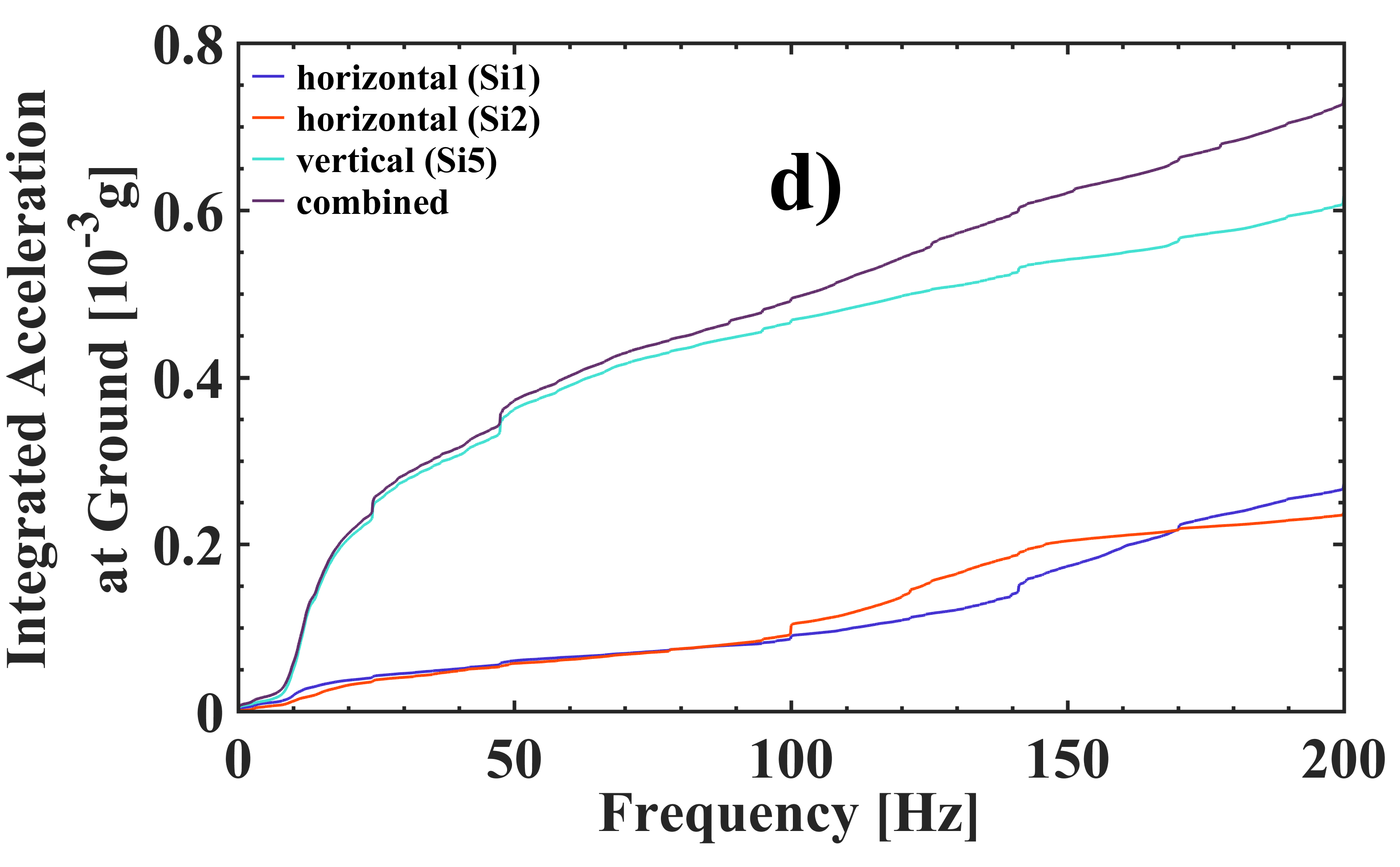}
	\par\end{centering}
	\caption{\label{fig:Cryocooler-Vibrations}Acceleration measured at three experimental
		plates inside the operating cryostat and ground acceleration. (a), (c) Spectral density of acceleration $S_{a,i}(f)$, $i=x,y,z$, in the three spatial directions. The red curve in panel (a) indicates the level measured on the plate that supports the 5~cm resonator described in this work. (b), (d) The total accelerations $\left[\int_{0}^{f}(S_{a,i}(f'))^{2}df'\right]^{1/2}$. The combined total acceleration is the root sum-of-squares of the
		three individual total accelerations.}
\end{figure*}

\section{Design}

\subsection{Cryostat accelerations\label{subsec: Cryostat-accelerations}}

The goal of our resonator design was to minimize its acceleration
sensitivity, in order to achieve good performance in closed-cycle
cryostats that provide no advanced vibration isolation. In such units,
the pulse-tube and seismic accelerations are transmitted to the optical
setup. As an example, our closed-cycle cryostat (Leiden Cryogenics, CF-1K) exhibits the acceleration
spectrum displayed in Fig.~\ref{fig:Cryocooler-Vibrations}~a. The
accelerations were measured on the optical setup, during operation
at $1.5$~K, for the three cartesian axes, using a precise interferometric
sensor. With an optical head of the sensor located outside the cryostat,
we used optical windows to reflect a laser beam from the device under
test. While it is best to measure the acceleration directly on the
resonator, here this was possible only for its vertical motion, because
of the limited optical access. To measure this motion, the laser interferometer
beam was reflected from one of its end faces. The motion along the
two horizontal directions was instead measured by reflecting the sensing
laser beams from two other silicon resonators (designated as "Si1" and "Si2") contained within the
cryostat.

The spectrum consists of peaks at multiples of $1.4$~Hz, the base
frequency of the pulse-tube cooler. The total acceleration integrated
over the frequency range from from $1$~Hz to $200$~Hz is $4.9\times10^{-3}$~$g$,
see Fig.~\ref{fig:Cryocooler-Vibrations}~b. The largest contribution
arises from the vibrations introduced by the rotary valve stepper
motor, which operates with a frequency close to $150$~Hz. Contribution
from the vibrations of the lab ground, measured with a high-sensitivity
piezoelectric transducer, is seen in Fig.~\ref{fig:Cryocooler-Vibrations}~c
and d, and is on the order of $0.9\times10^{-3}$~$g$. It is dominated
by the vertical component.

Because of this high acceleration level (compared to a standard room-temperature set-up
placed on a standard active vibration isolation platform) a minimized acceleration
sensitivity of the resonator frequency is clearly necessary.

\subsection{FEM Simulations}

 The resonator developed in this work is a vertically-oriented, axially
symmetric structure, supported at three points. It follows the concept
presented in refs.~\cite{Kessler2012a,Hagemann2014,Matei2017,Zhang2017,Robinson2019},
but was further simplified by avoiding the conical spacer shape and
employing instead a simple cylindrical shape. The spacer diameter
and length were chosen to be $37$~mm and $L=50$~mm, respectively,
the mirror substrates are of standard one-inch diameter and $6.3$~mm
thick. All the other geometry parameters were optimized using a commercial
finite-element-method (FEM) package (Ansys). The optical axis of the
resonator was aligned with the {[}111{]} crystallographic direction
of the silicon crystal, which is the direction with the highest Young's
modulus. We used the silicon stiffness matrix from Ref.~\onlinecite{Hall1967}
in our simulation.
\begin{figure*}[tbh]
	\begin{centering}
	     \includegraphics[width=0.48\textwidth]{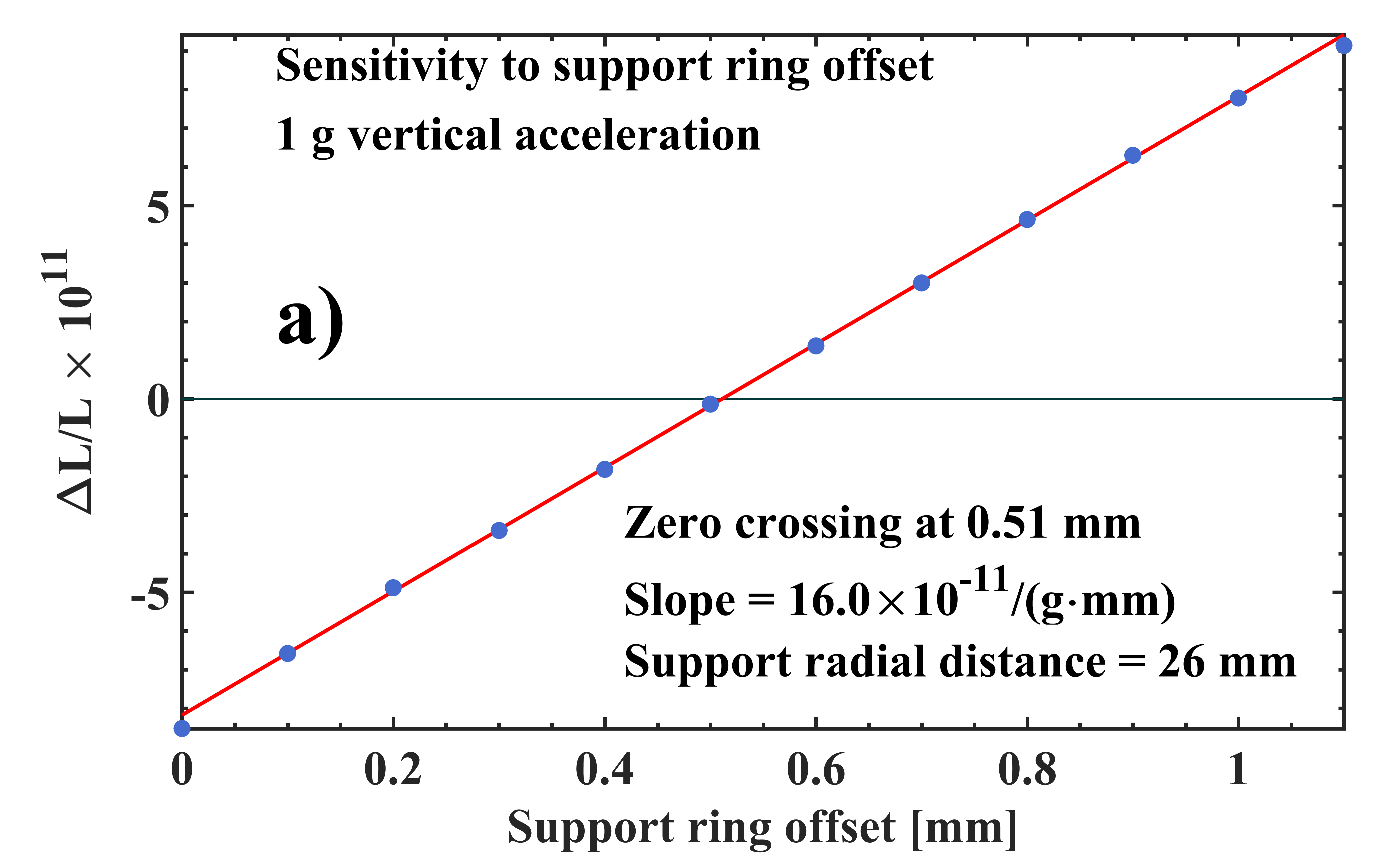}
		 \includegraphics[width=0.48\textwidth]{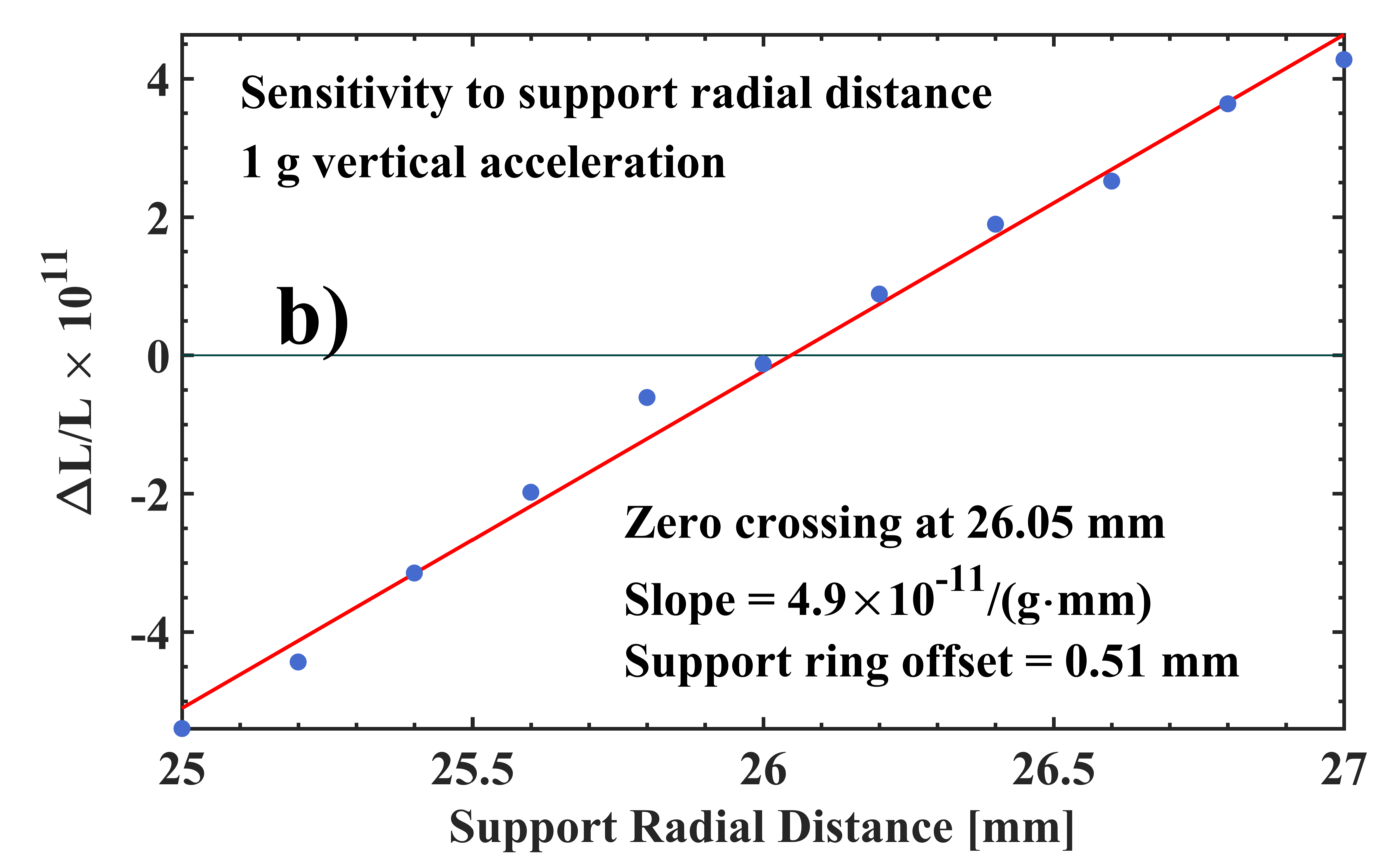}
		 \includegraphics[width=0.48\textwidth]{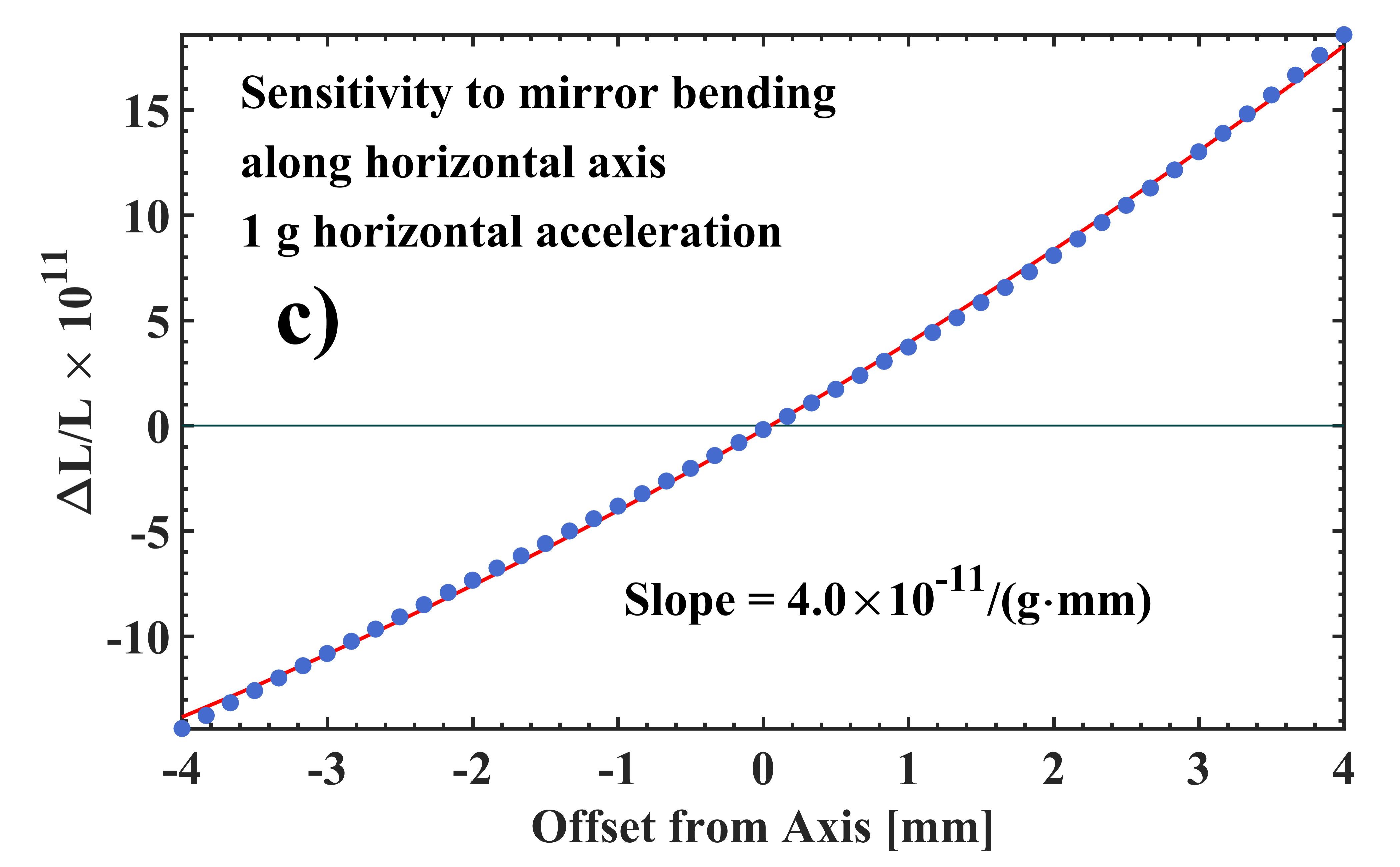}
		 \includegraphics[width=0.48\textwidth]{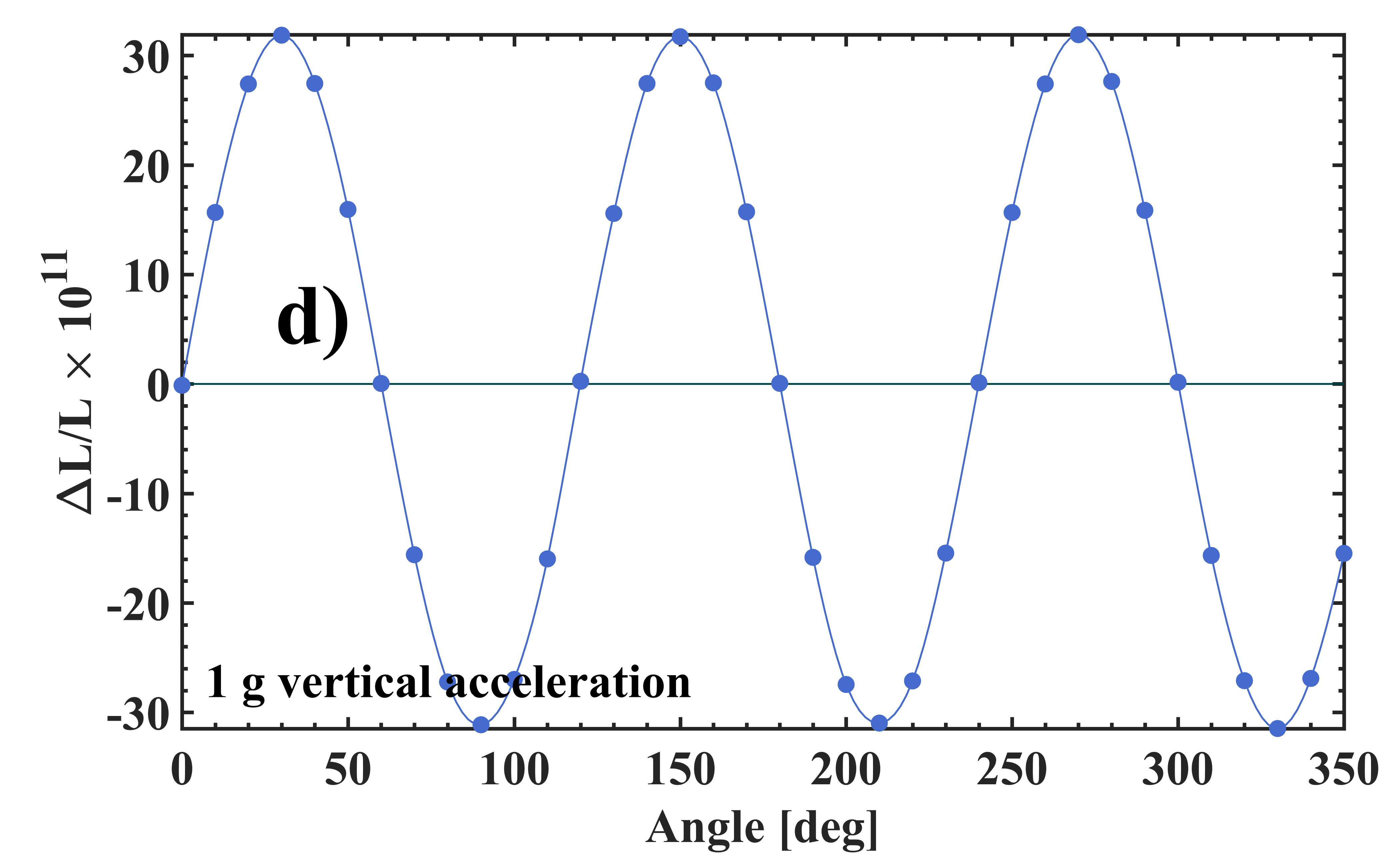}
		 \includegraphics[width=0.45\textwidth]{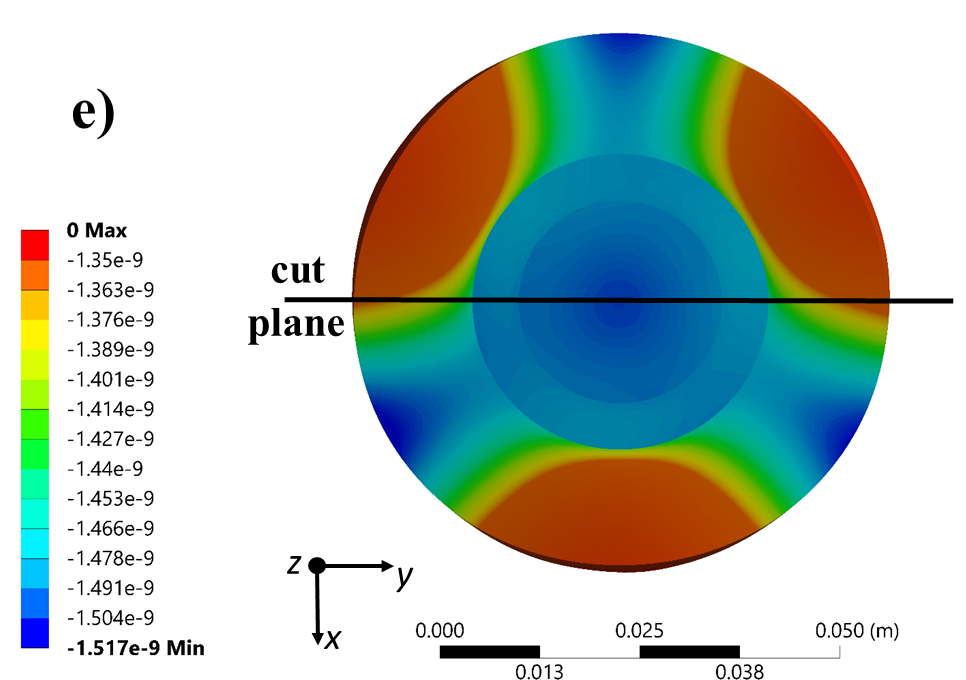}
		 \includegraphics[width=0.45\textwidth]{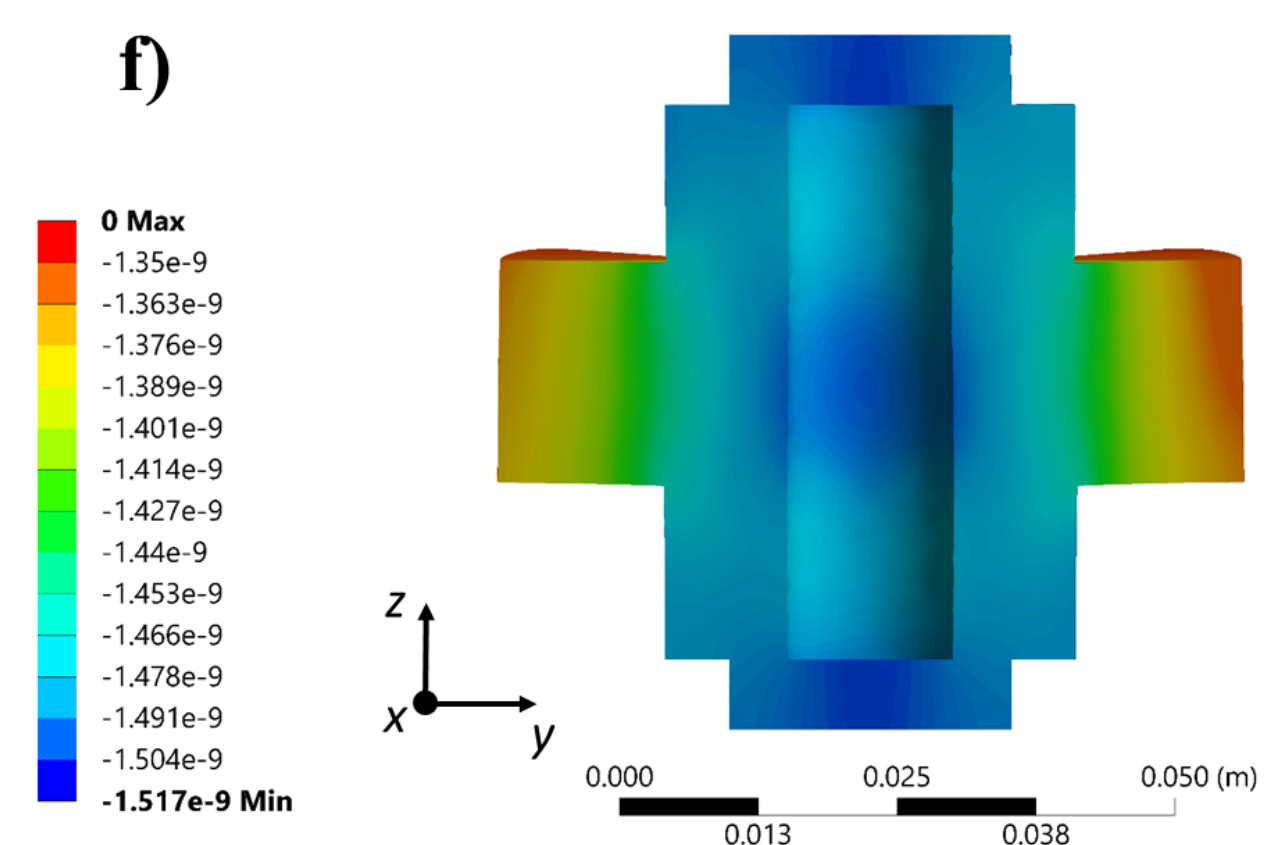}
		\par\end{centering}
	\caption{\label{fig:FEM-Results}Acceleration sensitivities of the resonator
		according to FEM simulations. (a) Influence of the imperfect manufacturing
		of the support ring on sensitivity during an application of vertical
		$1$~$g$ acceleration. (b) Sensitivity to changes in the radial
		position of the three supports and $1$~$g$ vertical acceleration.
		(c) Sensitivity to offset of the optical axis from the symmetry axis
		of the resonator ($1$~$g$ acceleration is applied in transverse
		direction). (d) Sensitivity to rotation of the resonator around the
		symmetry axis and an application of $1$~$g$ vertical acceleration.
		(e), (f) examples of the simulation results for the resonator with
		optimized shape. Color indicates the displacement in meter along the
		vertical direction (z-axis) due to the application of $1$~$g$ vertical
		acceleration. In (e) the displacement of the surfaces is shown; in
		(f) the displacements of the volume elements in the mid-plane cut
		along the vertical YZ plane (see definition of the cut plane in (e))
		are shown.}
\end{figure*}
The optimization was done by calculating the acceleration sensitivity
of the resonator, defined as fractional length change per unit acceleration,
$\Delta L/(a_{i}\,L)$, for different values of the geometrical parameters
and for different directions $i$ of the acceleration. After defining
a set of values we studied the influence of imprecise optical contacting,
of an offset of the resonator's position relative to the three support
points (fixed in space) and of manufacturing tolerance (assumed to
be $0.1$~mm). Our optimization was aimed at minimization of these
three sensitivities. This resulted in a support ring with diameter
and thickness of $67$~mm and $20$~mm, respectively, an offset
of the ring from the horizontal center plane toward the top of $0.51$~mm,
a diameter of the central bore of $15$~mm. The venting hole has
$2$~mm diameter, is located in the upper half of the resonator,
is centered $5$~mm below the top surface and forms an angle of $63$~deg
with the {[}100{]} crystallographic direction of silicon crystal.
Partial results of the sensitivity calculations are displayed in Fig.~\ref{fig:FEM-Results}~a
- c. With three supports placed at a radial distance of $26$~mm
from the optical axis of the resonator, the acceleration sensitivity
variation with vertical offset of the support ring, under $1$~$g$
vertical acceleration, is $16\times10^{-11}/\mathrm{(\mathit{g}\cdot mm)}$.
The sensitivity to an imperfect radial positioning of the resonator
relative to the supports is $5\times10^{-11}/(\mathrm{\mathit{g}\cdot mm)}$
and the sensitivity to a parallel offset of the optical axis from
the symmetry axis of the resonator is $4\times10^{-11}/(\mathrm{\mathit{g}\cdot mm)}$.

We also simulated the effect of rotation of the resonator around the
vertical symmetry axis, while the support points remained fixed in
space. The result is presented in Fig.~\ref{fig:FEM-Results}~d.
It resembles closely the result obtained by Matei et al. (Ref.~\onlinecite{Matei2016}).
The sensitivity is periodic with a period of $120$~deg, due to silicon's
anisotropic crystal structure, and has an amplitude of $3\times10^{-10}/\mathrm{\mathit{g}}$.
This results in a slope of $5\times10^{-12}/\mathrm{(\mathit{g}}\cdot\mathrm{deg)}$
around the sensitivity's zero crossing.

\begin{figure}[tbh]
	\begin{centering}
	        \includegraphics[width=0.48\textwidth]{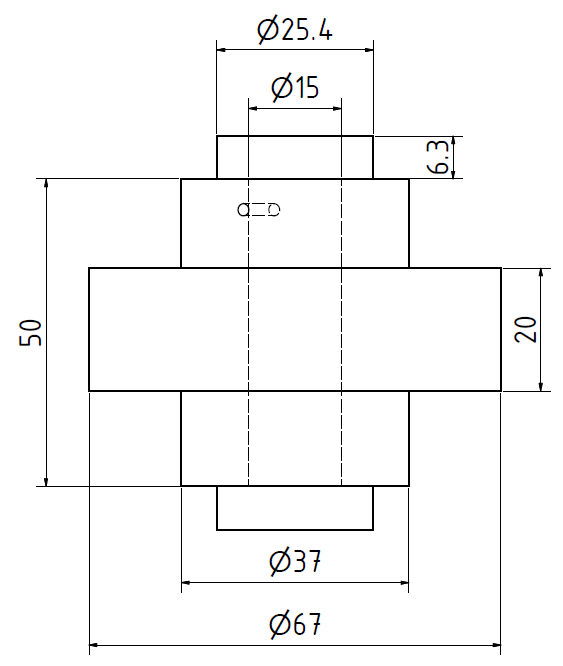}
		\par\end{centering}
	\caption{\label{fig:Si5-Design-Parameters} Optimized design of the resonator,
		as determined using FEM simulations. The $2$~mm diameter venting
		hole is located in the upper half at a distance of $5$~mm from the
		top end. Dimensions are in mm.}
\end{figure}

Fig.~\ref{fig:FEM-Results}~e and f display the deformation
of the resonator of optimum shape (without manufacturing errors) and
placed on the three supports at optimum position, with resultant zero
sensitivities. In the simulation, a vertically oriented $1$~$g$
acceleration is applied to all volume elements. The simulation reveals
a displacement of the top and the bottom mirrors by the same amount,
approximately $1.5$~nm, leaving the distance between them and thus the
resonator frequency unchanged. 
Final geometry of the resonator after FEM optimization is presented in Fig.~\ref{fig:Si5-Design-Parameters}.

\subsection{Modeling of thermal noise}

Random Brownian movement of the atoms in the crystal lattice of the
spacer and of the substrates as well as in the mirror coatings results
in a random fluctuation of the distance between the mirror internal
surfaces \cite{Numata2004a}. Thus, this motion directly influences
the optical path length and sets fundamental limits to the frequency
stability of a laser wave whose frequency is locked to the resonator.
For future reference, we calculate the thermal-noise induced instability
of the resonator length at $1.5$~K, assuming the parameters listed
in Tab.~\ref{tab:Thermal-Noise-Constants-1}. The results are presented
in Tab.~\ref{tab:Thermal-Noise}. A thirteen-fold reduction of thermal
noise from $300$~K to $1.5$~K is predicted, to the level $2.8\times10^{-17}$
at $1.5$~K. The coating contributes over $96$\% to the total noise,
because of its amorphous nature. The use of crystalline mirrors could
here provide a significant further reduction. In the present work
the performance of the system is not limited by the thermal noise.

\begin{table*}[tbh]
\begin{centering}
\begin{tabular}{c|c|c}
\hline 
\begin{minipage}[c][5mm][t]{0.1mm}%
\end{minipage}
Symbol & Parameter & Value\tabularnewline
\hline 
\hline 
\begin{minipage}[c][5mm][t]{0.1mm}%
\end{minipage}
$\lambda$ & laser light wavelength & $1562$~nm\tabularnewline
\hline 
\begin{minipage}[c][5mm][t]{0.1mm}%
\end{minipage}
$L$ & length of the spacer & $50$~mm\tabularnewline
\hline 
\begin{minipage}[c][5mm][t]{0.1mm}%
\end{minipage}
$R_{\mathrm{sp}}$ & radius of the spacer & $18.5$~mm\tabularnewline
\hline 
\begin{minipage}[c][5mm][t]{0.1mm}%
\end{minipage}
$r_{b}$ & radius of the central bore & $7.5$~mm\tabularnewline
\hline 
\begin{minipage}[c][5mm][t]{0.1mm}%
\end{minipage}
$w_{\mathrm{R=1m}}$ & beam waist at curved mirror & $338$~\textmu m\tabularnewline
\hline 
\begin{minipage}[c][5mm][t]{0.1mm}%
\end{minipage}
$w_{\mathrm{R=\infty}}$ & beam waist at flat mirror & $329$~\textmu m\tabularnewline
\hline 
\begin{minipage}[c][5mm][t]{0.1mm}%
\end{minipage}
$E$ & Young's modulus of Si along {[}111{]} crystallographic direction \cite{Zhang2014} & $187.9$~GPa\tabularnewline
\hline 
\begin{minipage}[c][5mm][t]{0.1mm}%
\end{minipage}
$\nu$ & Poisson's ratio of Si along {[}111{]} crystallographic direction \cite{Zhang2014} & $0.18$\tabularnewline
\hline 
\begin{minipage}[c][5mm][t]{0.1mm}%
\end{minipage}
$Q_{\mathrm{Si}}=1/\phi_{\mathrm{Si}}$ & Si quality factor \cite{Nawrodt2008} & $10^{8}$\tabularnewline
\hline 
\begin{minipage}[c][5mm][t]{0.1mm}%
\end{minipage}
$\phi_{\mathrm{ct}}$ & coating loss factor \cite{Granata2013} & $1$~mrad\tabularnewline
\hline 
\begin{minipage}[c][5mm][t]{0.1mm}%
\end{minipage}
$d_{\mathrm{ct}}$ & coating thickness & $9.38$~\textmu m\tabularnewline
\hline 
\end{tabular}
\par\end{centering}
\caption{\label{tab:Thermal-Noise-Constants-1}Parameters used for the calculation
of thermal noise.}
\end{table*}

\begin{table}[tbh]
\centering{}%
\begin{tabular}{c|c|c|c|c|c}
\hline 
\begin{minipage}[c][5mm][t]{0.1mm}%
\end{minipage}
 & \multicolumn{5}{c}{$\mathrm{\sigma_{y}\:(10^{-17})}$}\tabularnewline
\hline 
\hline 
\begin{minipage}[c][5mm][t]{0.1mm}%
\end{minipage}
$\mathrm{\mathbf{\mathrm{Temperature}}}$ & $\mathbf{\mathrm{300\:\mathrm{K}}}$ & $\mathbf{\mathrm{124\:\mathrm{K}}}$ & $\mathbf{\mathrm{16.8\:\mathrm{K}}}$ & $\mathbf{\mathrm{4\:\mathrm{K}}}$ & $\mathbf{\mathrm{1.5\:\mathrm{K}}}$\tabularnewline
\hline 
\begin{minipage}[c][5mm][t]{0.1mm}%
\end{minipage}
Spacer & $0.21$ & $0.13$ & $0.05$ & $0.02$ & $0.01$\tabularnewline
\hline 
\begin{minipage}[c][5mm][t]{0.1mm}%
\end{minipage}
Substrates & $1.31$ & $0.84$ & $0.3$ & $0.15$ & $0.09$\tabularnewline
\hline 
\begin{minipage}[c][5mm][t]{0.1mm}%
\end{minipage}
Coatings & $39.0$ & $25.1$ & $9.2$ & $4.5$ & $2.8$\tabularnewline
\hline 
\begin{minipage}[c][5mm][t]{0.1mm}%
\end{minipage}
$\mathrm{\mathbf{\mathrm{Total}}}$ & $39.0$ & $\mathbf{\mathrm{25.1}}$ & $\mathbf{\mathrm{9.2}}$ & $\mathrm{\mathbf{\mathrm{4.5}}}$ & $\mathbf{\mathrm{2.8}}$\tabularnewline
\hline 
\end{tabular}\caption{\label{tab:Thermal-Noise}Computed fractional frequency instability
of the mirror distance due to Brownian noise, for different operating
temperature. The contributions from the spacer, the substrates, and
the coatings are given. The parameters of Tab.~\ref{tab:Thermal-Noise-Constants-1}
were assumed. The fractional frequency instability is expressed as
an Allan deviation $\sigma_{\mathrm{y}}$, which is independent of
integration time.}
\end{table}

\section{Apparatus}

\subsection{Resonator and resonator support}

The resonator (denoted by "Si5" in some figures) was manufactured from a cylindrical silicon crystal
(resistivity $8$~kOhm/cm, diameter 4 inch), grown along the {[}111{]}
crystallographic direction using the float zone method. The optical
axis of the resonator is aligned with this direction. The two end
faces were polished to optical quality. High-reflectivity dielectric
mirrors for 1.5~$\mu$m wavelength were optically contacted to them
in-house. These silicon substrates originate from a different block
of material, having a resistivity of $4$~kOhm/cm and a diameter
of 4 inch. Their symmetry axes are oriented along the {[}100{]} crystallographic
direction. This aspect was not included in the above simulations.

\begin{figure*}[tbh]
\begin{centering}
\includegraphics[width=0.96\textwidth]{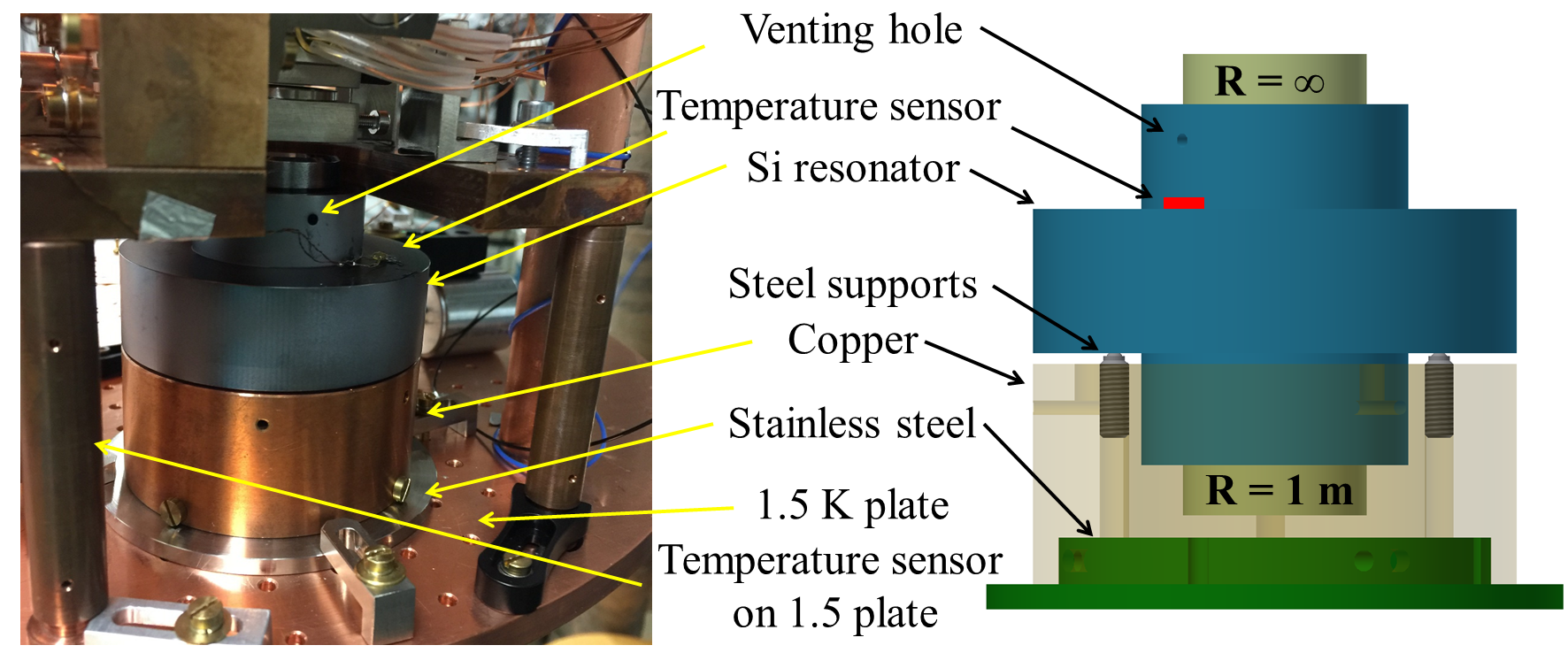}
\par\end{centering}
\caption{\label{fig:Support-Structure}Cryogenic resonator setup (left) and
corresponding schematic (right).}
\end{figure*}

The resonator was installed in an optical set-up inside a pulse-tube
cryostat equipped with a Joule-Thomson stage (Leiden Cryogenics).
A picture of the set-up and the corresponding schematic are shown
in Fig.~\ref{fig:Support-Structure}. The resonator was supported
from below at three points. The supports were pressure screws with
stainless steel balls at their ends. The balls were cut in half so
as to produce a circular surface of $3$~mm diameter. To increase
the friction between the balls and the resonator a layer of indium
foil was placed between them. In order to reduce the fluctuations
of the resonator's temperature caused by the fluctuations of cryostat
temperature, we split the support into two parts: a copper cylinder
as intermediate part and a stainless steel base. The latter acts as
a thermal low-pass filter, given its reduced thermal conductivity
compared to copper. The resonator temperature was measured by a sensor
(cernox) attached to the top surface of the silicon support ring using
a cryogenic grease.

\subsection{Cryogenic optical setup}

\begin{figure}[tbh]
\begin{centering}
\includegraphics[width=0.48\textwidth]{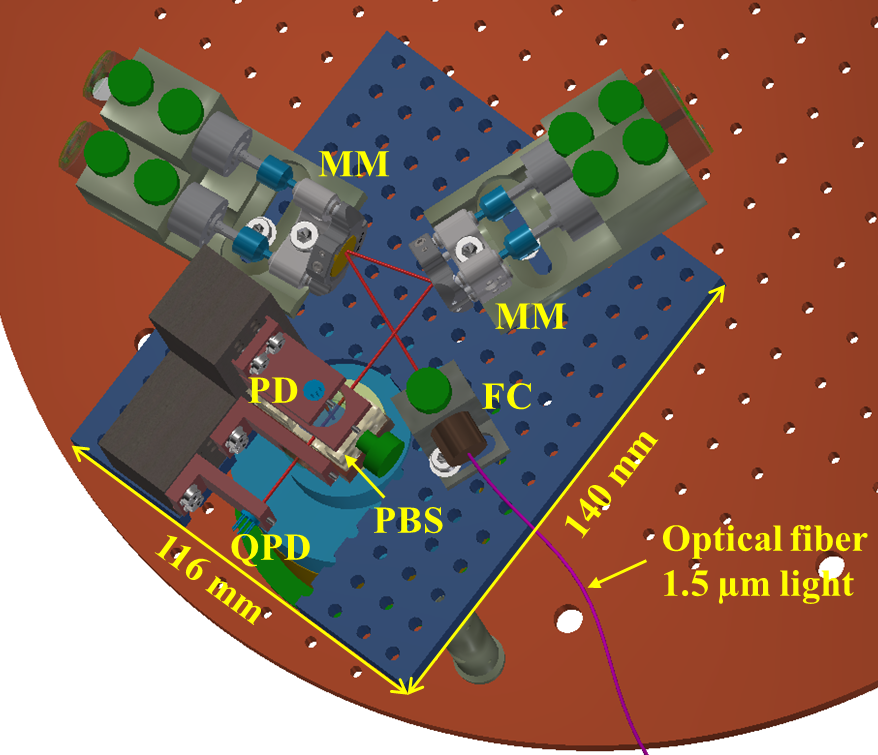}
\par\end{centering}
\caption{\label{fig: CAD Optical Breadboard} CAD figure of the cryogenic optical
setup. $\mathrm{FC}$, fiber collimator; $\mathrm{MM}$, motorized
mirrors; $\mathrm{PBS}$, polarizing beam splitter; $\mathrm{QPD}$,
quadrant photo detector; $\mathrm{PD}$, photodetector; $\mathrm{CAM},$
high-sensitivity infrared InGaAs camera. Red lines indicate free-space
paths.}
\end{figure}

The coupling of the laser light into the resonator and the detection
of the resonator response are performed on a cryogenic breadboard with a footprint of 116 mm x 140 mm,
shown in Fig.~\ref{fig: CAD Optical Breadboard}. The compact design minimizes
the optical path length in order to reduce the effect of unavoidable
misalignments upon cooling to cryogenic temperature. Additionally,
it incorporates two motorized mirror mounts allowing to correct for
the misalignments.

The light of the laser is carried to the breadboard setup using a single-mode
polarization-maintaining fiber. The end
of the fiber is fixed to the breadboard and coupled out using the
fiber collimator $\mathrm{FC}$. The wave is guided to the resonator
by reflecting off two motorized mirrors $\mathrm{MM}.$ Upon reaching
the polarizing beam splitter $\mathrm{PBS}$, a small part of the
light is diverted to the quadrant photodetector, marked as $\mathrm{QPD}$,
for monitoring of the beam position. The remainder is guided to the
resonator, passing through a quarter-wave plate (not shown in the
schematic). It is partially reflected from the front mirror. The reflected
light is detected by a high-bandwidth photodetector $\mathrm{PD}$.
The signal can be used for a Pound-Drever-Hall-type lock ($\mathrm{PDH}$),
not implemented here. The light transmitted by the resonator is split by a beam splitter. One part is detected by
a photodiode installed below the experimental plate (not shown). The
other part exits the cryostat through a window and is used for identification
of the transverse mode excited by the laser. For this purpose, a high-sensitivity
room-temperature InGaAs camera is installed outside the cryostat in
the beam path.

\subsection{Optical frequency stabilization system and procedures}

\begin{figure*}[tbh]
\begin{centering}
\includegraphics[width=0.96\textwidth]{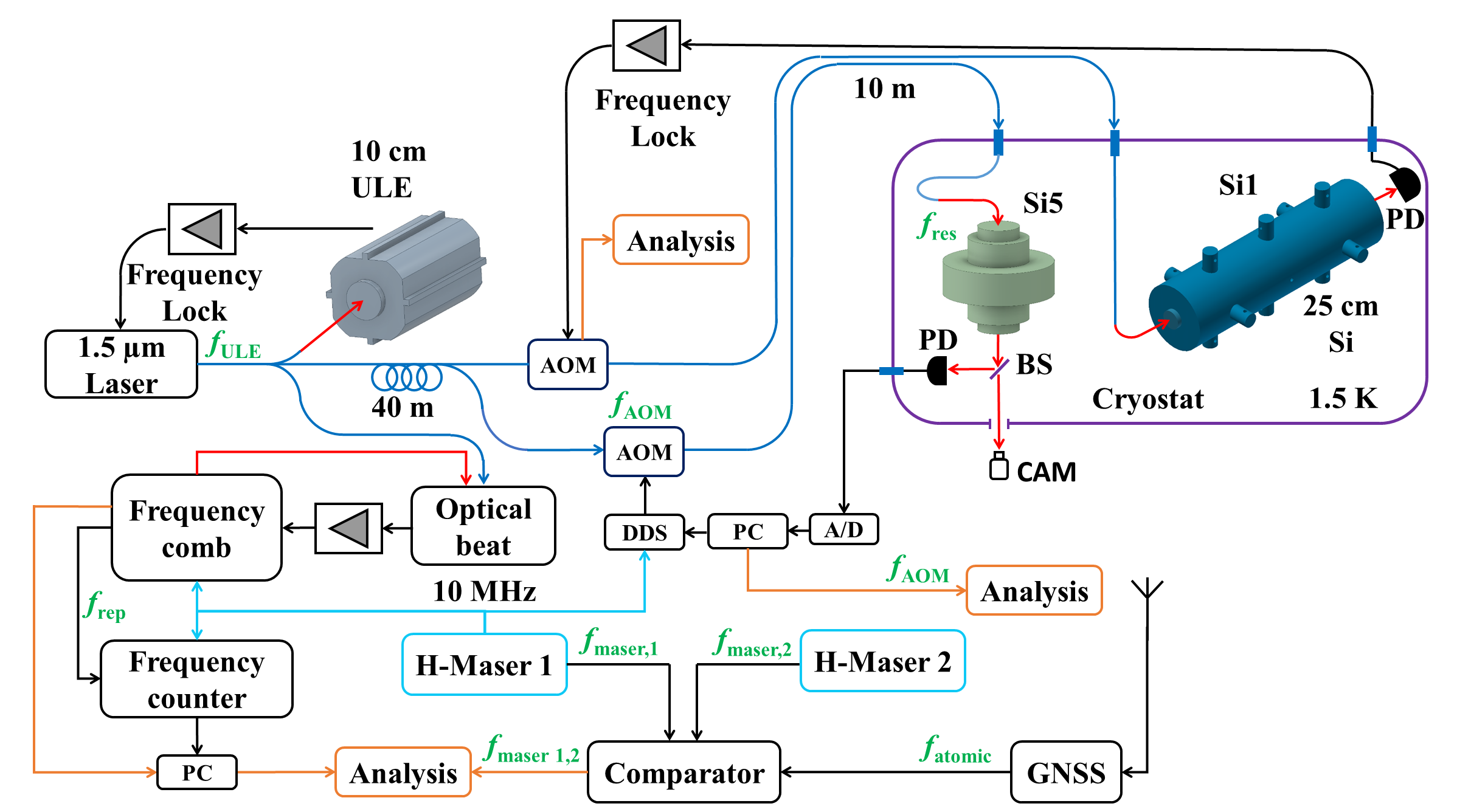}
\par\end{centering}
\caption{\label{fig:Schematic-Optical-Setup} Overall experimental setup. Red
and blue lines indicate free-space and fiber-coupled optical paths, respectively.
Black lines represent the electronic paths and the light blue line
is the $10$~MHz reference signal from the hydrogen maser. $f_{\mathrm{res}}$,
frequency of the cryogenic silicon resonator; $f_{\mathrm{ULE}}$,
frequency of the room-temperature ULE resonator; $f_{\mathrm{atomic}}$,
frequency of the GNSS satellite signal; $f_{\mathrm{maser,i}}$, frequencies
of two hydrogen masers; $f_{\mathrm{rep}}$, repetition rate of the
frequency comb; $f_{\mathrm{AOM}}$, frequency of the AOM; PC, personal
computer; AOM, acousto-optical modulator; DDS, direct digital synthesizer;
A/D, analog to digital signal converter; PD, photodiode; BS, beam
splitter; CAM, InGaAs infrared camera for mode detection.}
\end{figure*}

The concept of the present system is to combine a laser with a good
short-term ($\tau\leq20$~s) frequency stability with a cryogenic resonator whose task
is to provide a better frequency stability on medium (> 20~s)
and long  (> 1000~s) time scales than possible with a room-temperature reference
resonator. Therefore, we use a room-temperature, 10~cm long ULE resonator
for the pre-stabilization of the laser's frequency. The pre-stabilized
laser then interrogates the $\mathrm{TEM_{00}}$ mode of the cryogenic
resonator. The overall layout of the optical setup is presented in
Fig.~\ref{fig:Schematic-Optical-Setup}. It includes components necessary
for the characterization of the system performance.

The pre-stabilized laser (optical frequency $f_{{\rm ULE}}$) has
a linewidth of less then $1$~Hz at its output. Part of this light
is transferred to the cryogenics lab and into the cryostat via an
approximatly $50$~m long optical fiber. Initially, no active fiber
noise cancellation was installed for the path between the pre-stabilized
laser and the cryogenic lab. The typical broadening of the linewidth
due to fiber noise along this $40$~m long path was measured
to be $20$~Hz. Additional noise is likely introduced by the vibrations
inside the cryostat along the remaining $10$~m long fiber, but could
not be measured independently because no wave reflected back from
the resonator or from the end of the fiber could be observed. One
part of the prestabilized laser light is used for measuring the laser
frequency $f_{\mathrm{ULE}}$ with a frequency comb referenced to
an active hydrogen maser. The maser is continuously compared to a
GNSS signal providing a reference frequency $f_{\mathrm{atomic}}$.
To improve the sensitivity of the frequency measurements we reduce
the spectral width of the comb lines by phase-locking the comb to
the pre-stabilized laser. The repetition rate $f_{\mathrm{rep}}$
of the comb is measured with a low-noise frequency counter and from
the data $f_{\mathrm{ULE}}/f_{\mathrm{maser}}$ is computed.

Another part of laser light, fiber-splitted in the cryogenic laboratory, is guided to the 
$25$~cm cryogenic ultra-low drift silicon resonator \cite{Wiens2016} (denoted by "Si1" in some figures) installed 
on another optical breadboard inside the same cryostat. This resonator is also employed  
as a reference for determination of the frequency instability of the $5$~cm resonator. 

We use an acousto-optic modulator (AOM with frequency $f_{{\rm AOM}}$)
to bridge the gap between the laser frequency $f_{\mathrm{ULE}}$ and the frequency
of the closest TEM\textsubscript{00} mode of the silicon resonator,
driven by a direct digital synthesizer (DDS) controlled by a personal
computer (PC).

To determine the frequency of the resonator we repeatedly measure
the line center frequency. Two techniques have been employed: (1)
scanning over the resonance line and (2) alternating interrogation
of the half-transmission points of the resonance. Both techniques
are compatible with the use of very low light power ($\leq$1~$\mu$W).
Also, they do not require continuous coupling of laser light into
the resonator, but can be applied, if desired, with low duty cycle.
Together, these features help to reduce permanent or semi-permanent
changes in the mirror coatings due to exposure to laser light.
Another advantage is the absence of offsets introduced by active optical
elements, e.g. residual amplitude modulation introduced by an electro-optical
modulator (EOM) or laboratory-temperature-induced variations of the
PDH electronics lock point. In the linescan technique, each center
frequency determination is carried out by sweeping the frequency of the laser light over the resonance
line with an AOM. The light transmitted through
the resonator was detected by the cryogenic detector and the signal
sampled by a $14$-bit DAQ card with $40$~kS/s. The frequency span
was set to twice the linewidth, $2\Delta\nu=40$~kHz, and the
(one-way) scan time was set to $0.7$~s. The data of two subsequent
scans, upwards and downwards in frequency, was averaged. This data
was subsequently fitted with a Lorentzian function to determine the
AOM frequency $f_{\mathrm{AOM}}$ corresponding to the resonator's
center frequency. This frequency value is thus obtained essentially
immediately after each pair of scans. A digital control modifies the
scan range settings so as to maintain the resonance frequency in the
center of the range. In addition, the value $f_{\mathrm{AOM}}$ is
stored and used to compute the resonator frequency as $f_{\mathrm{res}}/f_{\mathrm{maser}}=f_{\mathrm{ULE}}/f_{\mathrm{maser}}+f_{\mathrm{AOM}}/f_{\mathrm{maser}}$.
In our measurements, the procedure is repeated continuously, but as
mentioned, a wait time interval could be inserted if needed.

The detuning technique is realized by periodically shifting the laser
light frequency by $\pm\Delta\nu/2$ relative the resonance frequency
$f_{res}$, to one of the two half-transmission detunings. The signal
of the transmission photodiode for $-\Delta\nu/2$ detuning is measured
at time step $i-1$ and averaged over a $500$~ms time period, yielding
the value $A_{i-1}$. Subsequently, the other half-transmission position
is selected and the corresponding amplitude $A_{i}$ recorded. Then,
the frequency correction $\Delta f_{\mathrm{AOM},i}=(\Delta\nu/2)(A_{i}-A_{i-1})/(A_{i}+A_{i+1})$\textcolor{red}{{}
}is calculated and applied to the AOM. The absolute frequency of the
resonator is computed as in the case of the linescan technique.

\section{Characterization}

\subsection{Temperature stability}

\begin{figure*}[ht]
	\begin{centering}
	 \includegraphics[width=0.48\textwidth]{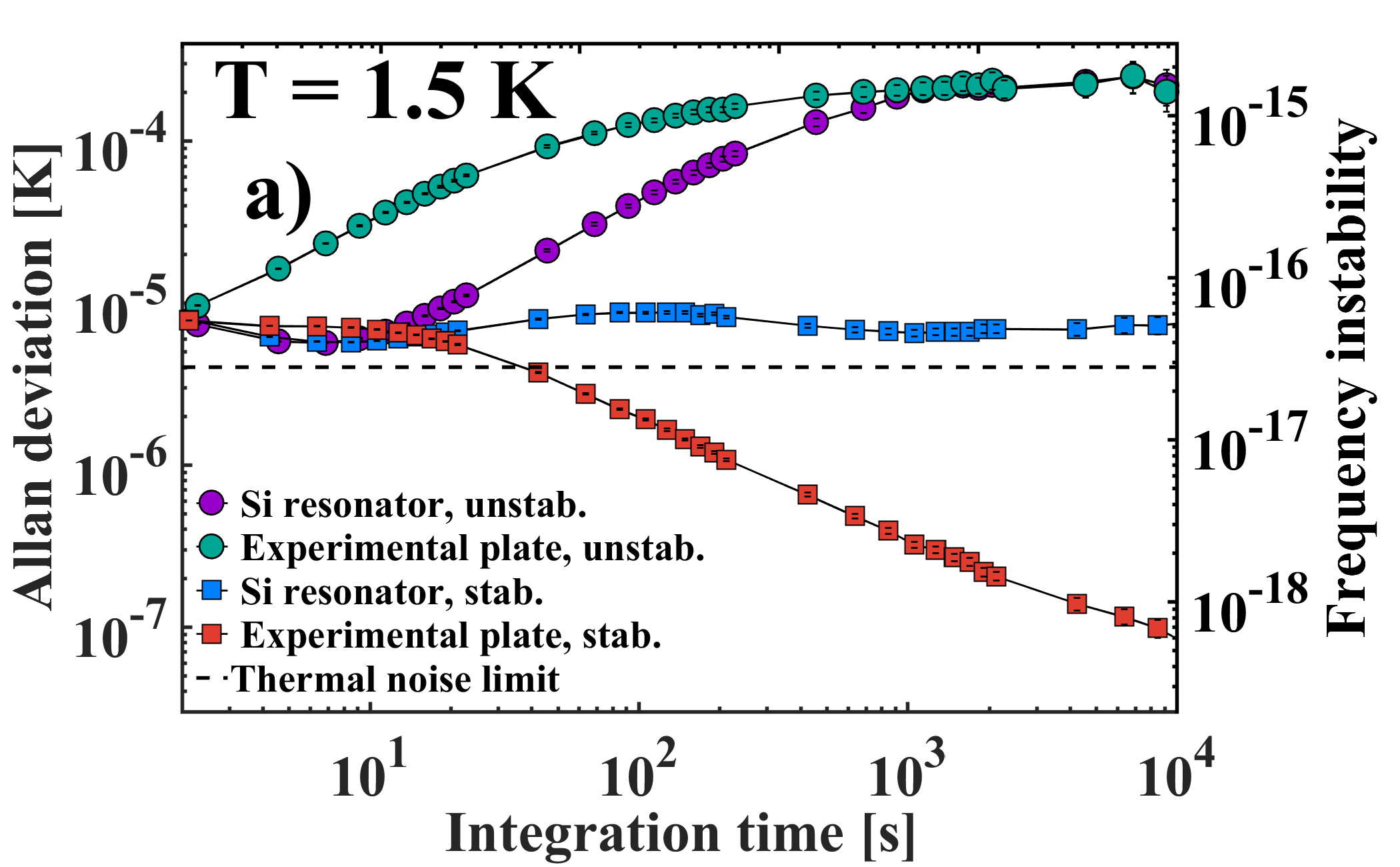}
	 \includegraphics[width=0.48\textwidth]{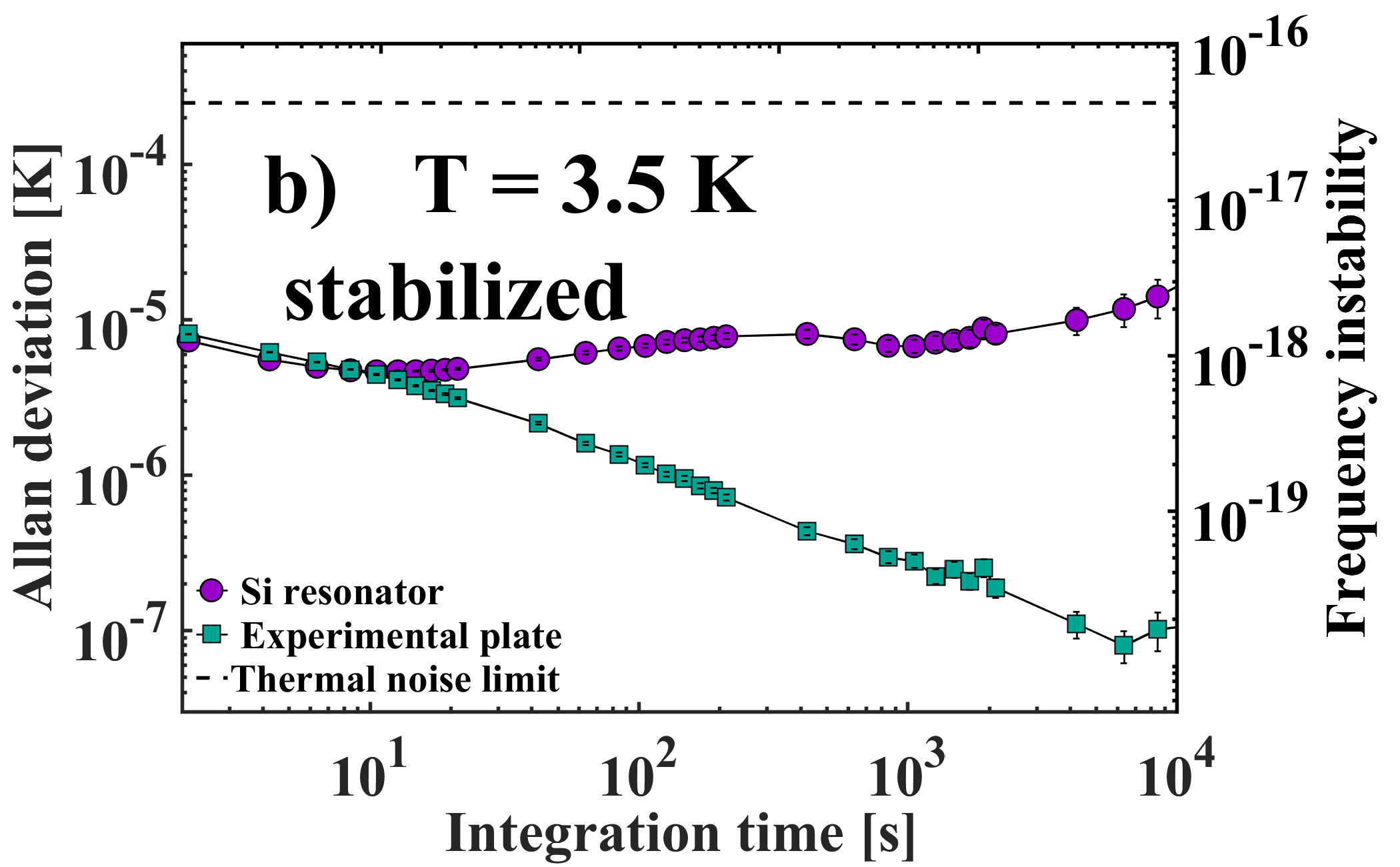}
	\par\end{centering}
\caption{\label{fig:Temperature-Instability}Temperature instability of the
experimental plate, measured with a sensor attached to it in close
proximity to the resonator support structure, and of the resonator
itself. (a) At $1.5$~K, with and without active temperature stabilization,
and (b) at $3.5$~K, with active stabilization. The right axis in
both diagrams shows the resulting fractional frequency instability,
computed using the appropriate CTFs. }
\end{figure*}

The temperature stability of the resonator is important since it can
affect its medium- and long-term frequency stability through the coefficient
of thermal sensitivity of frequency (CTF). Fig.~\ref{fig:Temperature-Instability}
compares the temperature instability measured on the resonator and
on the base plate, at two operating temperatures, and under two different
operating modes. The laser light was blocked during these measurements.
The temperature instability is given as a modified Allan deviation
and is computed from the temperature time series. At the temperature
of $1.5$~K the instability of the free-running resonator temperature
is lower than that of the base plate for integration times up to $1000$~s.
This shows that the baseplate temperature variations are substantially
attenuated by the resonator support structure. The smallest temperature
instability is $4.5$~$\text{\textmu}$K at $6$~s integration time.
The instability data combined with the thermal expansion coefficient
of the resonator at 1.5~K, $7\times10^{-12}$/K, yields an estimate
of the frequency instability of the resonator. This is shown on the
right axis of Fig.~\ref{fig:Temperature-Instability}. The instability
is above the level of the calculated thermal noise for all integration
times. Thus, for operation at 1.5~K the current support structure
would represent a limiting factor for the future resonator performance.
An improvement could be realized by  introducing a second stage of passive thermal
isolation or, as done in this work, by actively stabilizing the temperature
of the base plate (see Fig.~\ref{fig:Temperature-Instability} a). This improvement was not necessary for the experimental results presented
here.

To operate the setup at $3.5$~K, a temperature of particular interest
(see below), an active temperature stabilization was implemented.
The resulting temperature instability of the resonator is $10$~$\text{\textmu}$K
or less for all integration times up to $10\,000$~s. The inferred
frequency instability of the resonator, depicted on the right axis,
was calculated assuming that the temperature set point has an undesired
offset of 0.02~K from the optimum temperature. Such a small deviation
is conservative, in view of the accuracy with which the optimum temperature
is in principle measurable, see Fig.~\ref{fig:Si5-CTE}~d. The absolute
value of the CTF is then less than $2\times10^{-13}/{\rm K}$. This
yields a frequency instability of less than $3\times10^{-18}$, significantly
below the thermal noise limit.

The temperature instability presented in Fig.~\ref{fig:Temperature-Instability}
must be viewed with caution. Variations of laboratory temperature
affect the reference voltage in the control electronics of the active
temperature stabilization circuit. The voltage variations are interpreted
as variations of the cryogenic temperature. This explains the large
difference between the instability measured by loop base plate sensor
and by the sensor attached to the resonator, seen in Fig.~\ref{fig:Temperature-Instability}
b. While the in-loop sensor suppresses the influence of the lab temperature
variations, the monitor sensor at the resonator does not. Therefore,
the temperature instability of the resonator is likely to be equal
or below the instability presented in Fig.~\ref{fig:Temperature-Instability}.

\subsection{Resonator properties}

\begin{figure*}[tbh]
	\begin{centering}
		\includegraphics[width=0.45\textwidth]{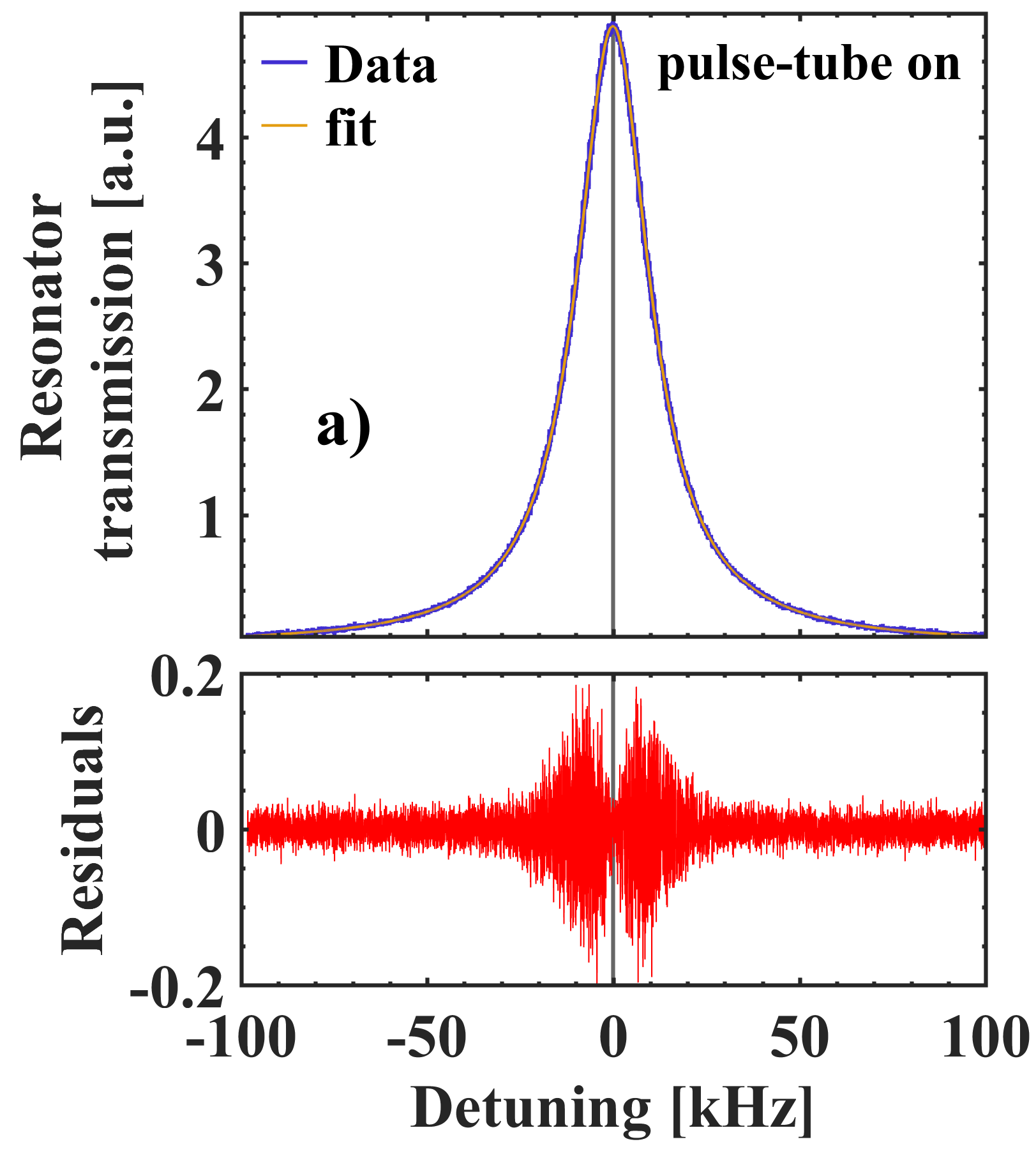}
		\includegraphics[width=0.45\textwidth]{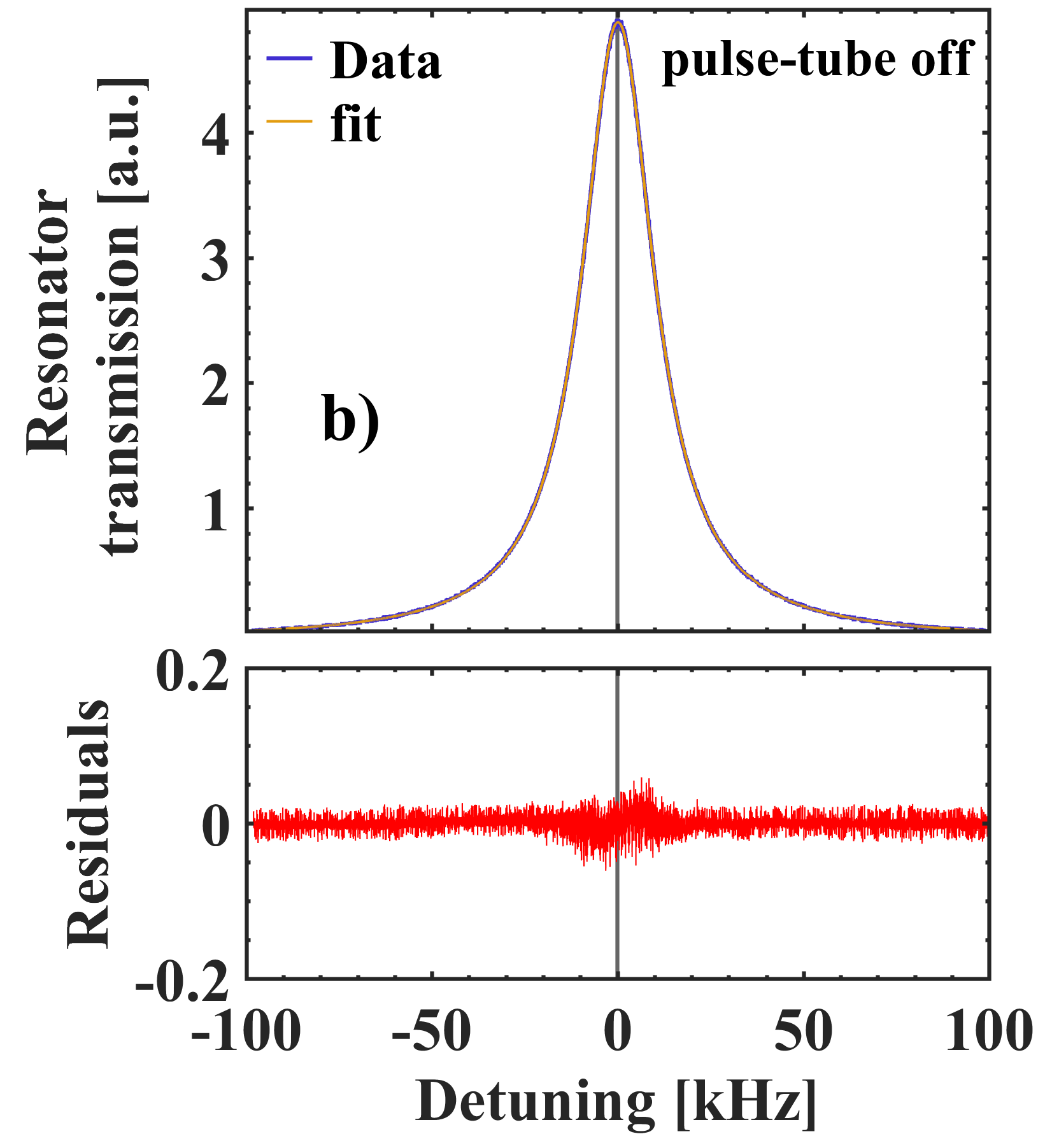}
		\par\end{centering}
\caption{\label{fig:Linewidth} One-way frequency scans of the pre-stabilized
laser frequency over the resonator TEM\protect\textsubscript{00}
mode. Scan duration is 80~s. The signal is the light power transmitted
through the resonator. The pulse-tube cooler is on (panel (a)) and
off (panel (b)). The orange lines are Lorentzian fits. Operating temperature:
$1.5$~K.}
\end{figure*}

After cooling down to $4$~K, only a slight optimization of the in-coupling
by actuating the mirrors was required. The cooling was then continued
down to the operational temperature of $1.5$~K. The TEM\textsubscript{00}
resonance was identified and could be routinely interrogated with
the pre-stabilized laser. A typical scan is presented in Fig.~\ref{fig:Linewidth}
a. From fits of Lorentzian functions to a series of scans we find
a mean full width at half maximum of $24.2\pm0.2$~kHz. This corresponds
to a finesse of $120\,000$. The measured incoupling $(1-P_{{\rm r,on}}/P_{{\rm r,off}})$,
where $P_{{\rm r,on}}$ is the on-resonance reflected power and $P_{{\rm r,off}}$
is the off-resonance reflected power, was 10\%. The mode matching
efficiency, determined by characterizing the incoupling
of other transverse modes, was $60$\%. The fit residuals in the figure deviate
clearly from near zero when the laser frequency is tuned to the vicinity
of the half-maximum resonator transmission frequency (bottom panel
in Fig.~\ref{fig:Linewidth}~a). This indicates the presence of
frequency fluctuations of the resonator. They are due to the vibrations
generated by the cryostat, as can be seen from the fact that the residuals are
weaker when the cooler is off (see Fig.~\ref{fig:Linewidth}~b).

\subsection{Acceleration sensitivity}

One of the goals of this work was minimization of the resonator's
acceleration sensitivity. The characterization of this property is
therefore of importance. It is usually done by shaking the resonator
along one of the three orthogonal spatial directions and observing
the frequency shift of the resonator. However, in our case, the structure
of the crystat prevented us from producing controlled vibrations in
desired directions, and thus made it impossible to measure the vibration
sensitivity for the three spatial directions individually. However,
we were able to estimate the overall sensitivity of the resonator
submitted to the accelerations produced by the cryostat by measuring
the variations of the transmission signal caused by the deformations
of the resonator.

\begin{figure*}[tbh]
	\begin{centering}
		\includegraphics[width=0.48\textwidth]{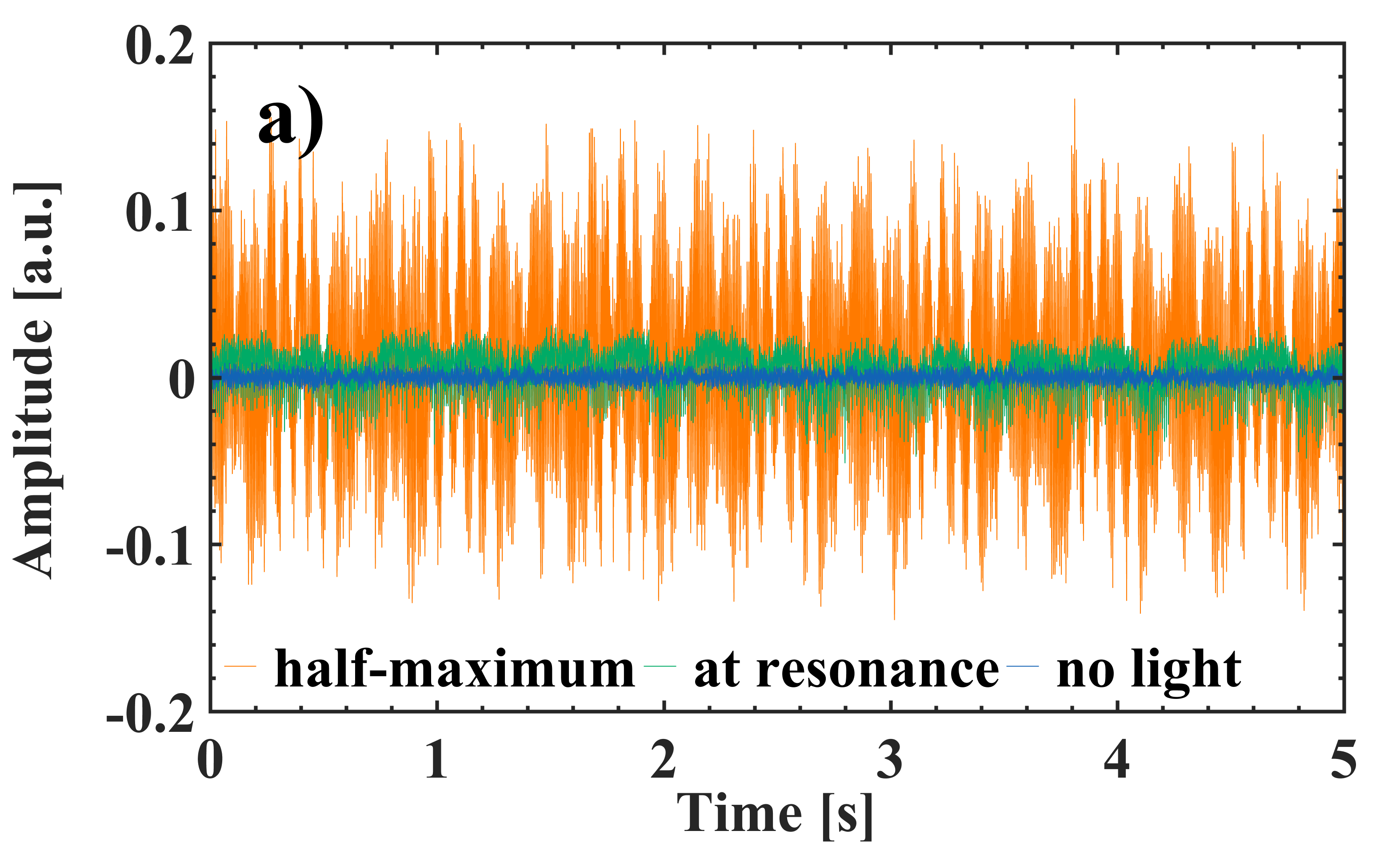}
		\includegraphics[width=0.48\textwidth]{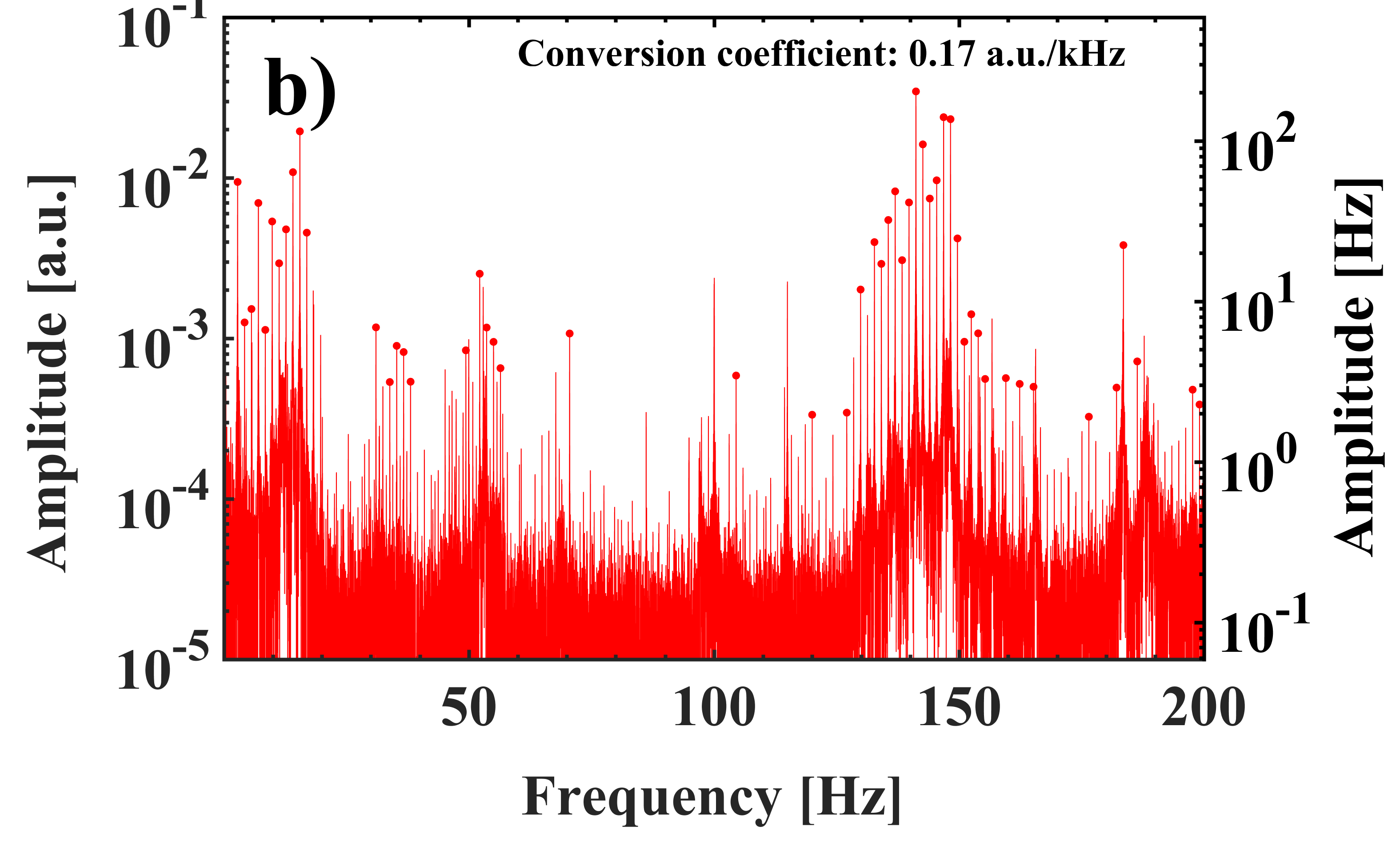}
		\includegraphics[width=0.48\textwidth]{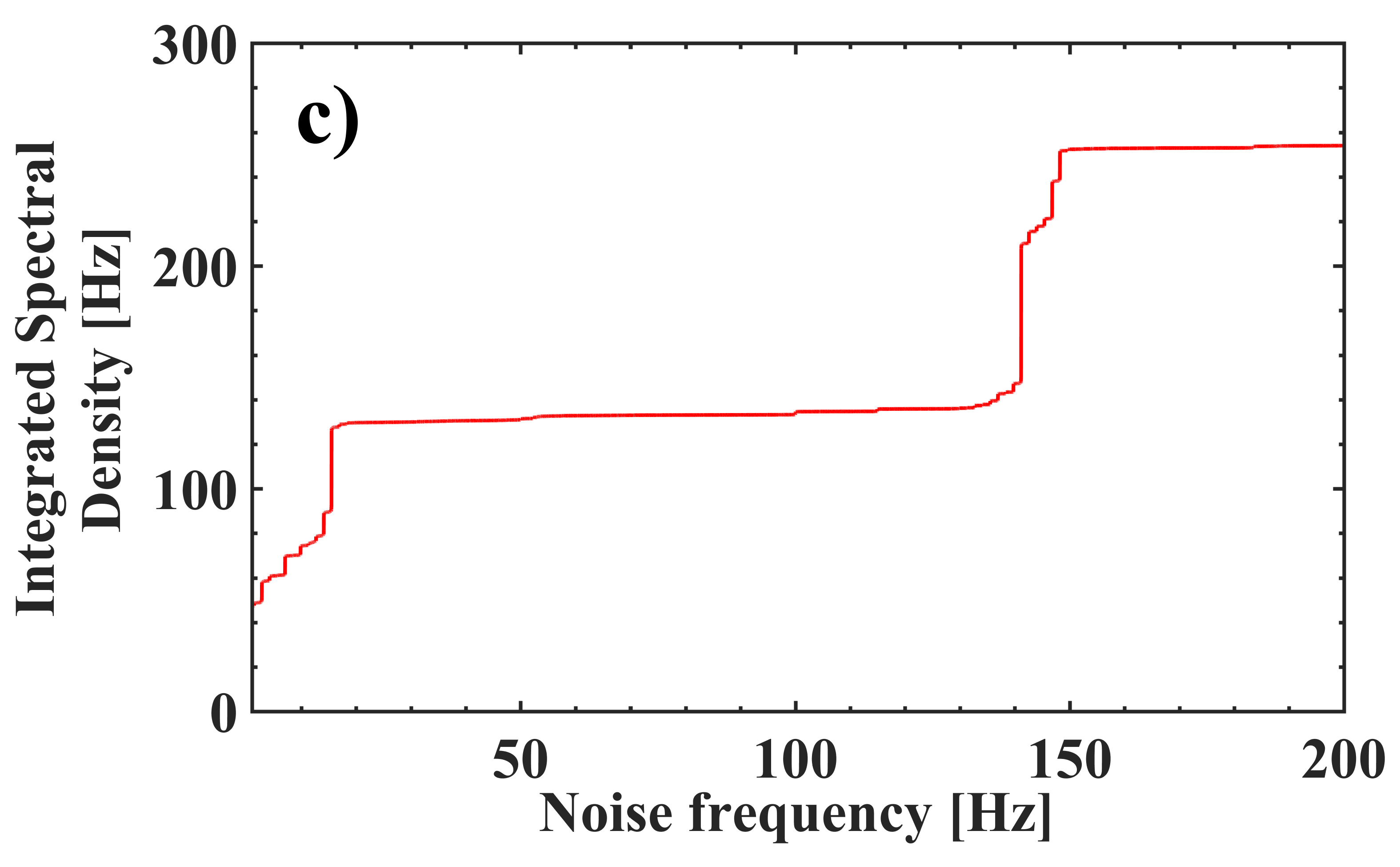}
		\includegraphics[width=0.48\textwidth]{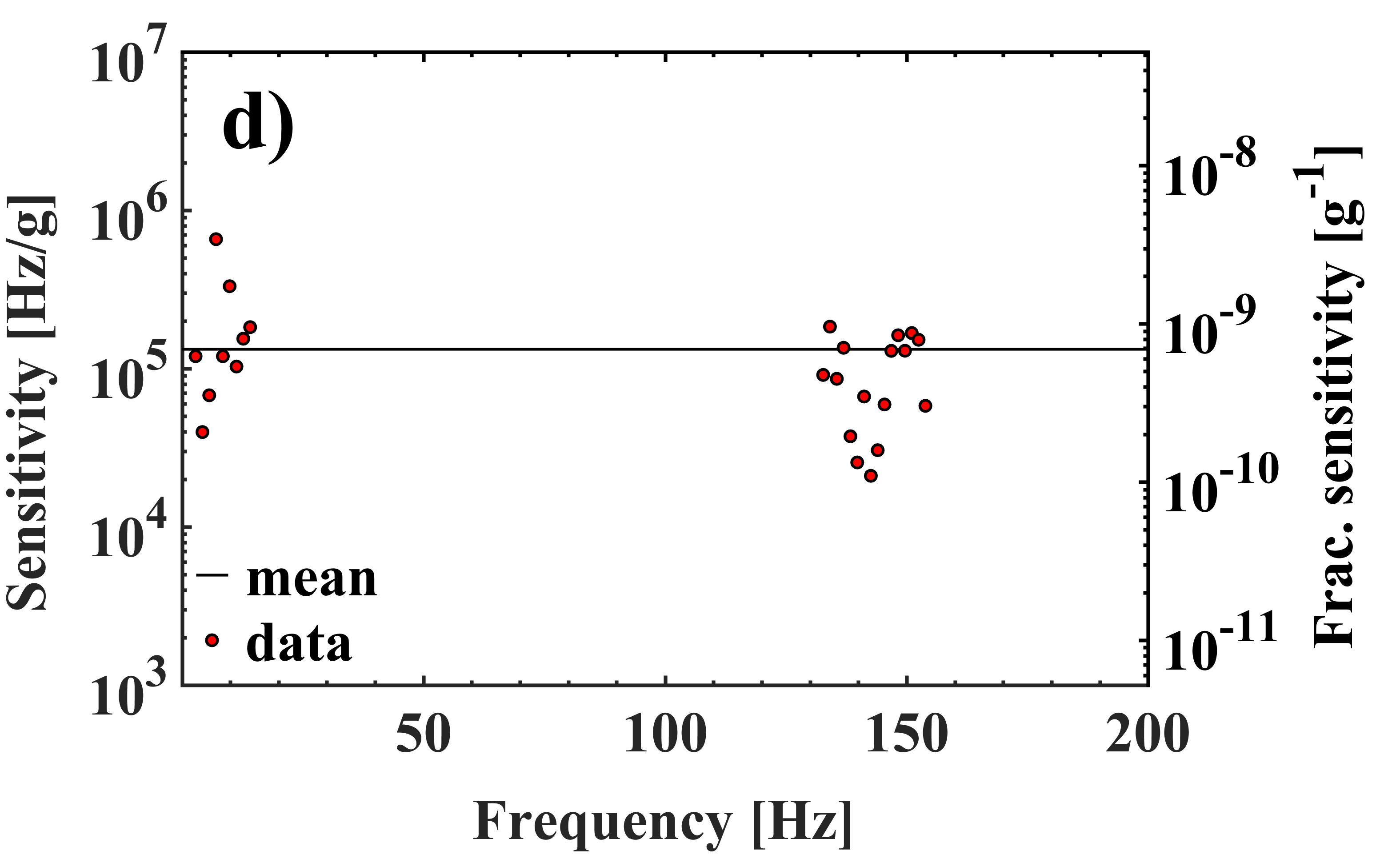}
		\par\end{centering}
\caption{\label{fig:Si5-Frequency-Sensitivity}Determination of the resonator
sensitivity to vibrations. (a) Time traces of the cavity transmission
signal, with subtracted offset, measured with the laser frequency
tuned to the half-transmission of the resonance (orange) and on-resonance
(green). The green time trace includes a factor 1/2 to account for
the larger transmitted power. Blue: background signal taken with laser
off. (b) Spectrum calculated from the time trace at half-maximum and
after subtraction of the spectrum taken with the laser tuned to the
resonance, and of the noise spectrum . Right y-axis gives the frequency
fluctuation level, after applying the conversion factor $S=0.17$~a.u./kHz
(slope of the transmission signal at half-maximum value). (c) Contribution
to the linewidth from vibrations in the frequency region up to $200$~Hz.
(d) Sensitivity of the resonator to vibrations at different frequencies
obtained by division of the spectrum from panel (b) by the spectrum of cryostat
accelerations defined as root sum-of-squares of the three individual accelerations. The black line indicates the mean of the shown data
points.}
\end{figure*}

An estimate of the sensitivity can be obtained using a side-of-fringe discriminator
technique. For this, the laser light is frequency-tuned so that the
(time-averaged) transmission signal is half the maximum. Then, the
fluctuations of the transmission signal are recorded. Fig.~\ref{fig:Si5-Frequency-Sensitivity}~a
shows a time trace (orange). For comparison, we also determined the
contribution arising from power variations of the laser wave. They
were measured with the laser frequency tuned to maximum transmission
of the resonator. Finally, we measured the background noise in the
coaxial cable by blocking the laser light. We calculated the r.m.s.
amplitude deviations corresponding to these three situations, $\Delta A{}_{\mathrm{S}}$,
$\Delta A{}_{\mathrm{T}}$, $\Delta A_{\mathrm{N}}$. The r.m.s. signal deviation
at the half-transmission detuning due to relative frequency fluctuations
between laser and resonator, $\Delta A_{\mathrm{eff}}$, is obtained
after subtraction of $\Delta A{}_{\mathrm{T}}/2+\Delta A_{\mathrm{N}}$ from $\Delta A{}_{\mathrm{S}}$. The correction compared
to using only $\Delta A_{S}$ is at the 1\% level.

Using $\Delta A_{\mathrm{eff}}$ together with the slope of resonator
transmission at half-maximum, $S=0.17$~a.u./kHz, and the total acceleration
$a_{\mathrm{total}}$, we obtain the acceleration sensitivity $\sigma_{\mathrm{total}}=61.5$~kHz/$g$
($3.2\times10^{-10}$/$g$). 

We obtained the spectrum of frequency fluctuations from the power
spectra corresponding to the time traces in Fig.~\ref{fig:Si5-Frequency-Sensitivity}~a.
It is presented in Fig.~\ref{fig:Si5-Frequency-Sensitivity}~b.
Conversion into frequency units was done using the transmission signal
sensitivity $S$. Integration of the power spectral density provides
an estimate of the linewidth that a laser locked to the resonator
would exhibit, assuming that the measured frequency fluctuations are
solely due to the resonator length fluctuations. Fig.~\ref{fig:Si5-Frequency-Sensitivity}~c
shows the integrated frequency noise. The total contribution within
the detected bandwidth is $250$~Hz. This value is in agreement with
the first estimate above. The frequency fluctuations, and thus the
acceleration noise in the frequency bands {[}$1$, $20${]}~Hz and
{[}$140$, $150${]}~Hz contribute most to the linewidth.

The data in Fig.~\ref{fig:Si5-Frequency-Sensitivity}~b together
with the acceleration data of Fig.~\ref{fig:Cryocooler-Vibrations}
allows a direct determination of the resonator acceleration sensitivity
as a function of frequency, by computing their ratio. We show this
for the frequency bands mentioned above in Fig.~\ref{fig:Si5-Frequency-Sensitivity}~d.
The mean sensitivity is $133$~kHz/$g$ ($6.9\times10^{-10}$/$g$).
Note that the individual values in the spectrum vary by one order
of magnitude around the mean. The largest value is $7\times10^{2}$~kHz/$g$
($5\times10^{-9}$/$g$) at a frequency of $7$~Hz. In view of the
fact that we cannot separate the contributions of the three spatial
components of the acceleration in the transmission signal, our analysis
cannot be more precise.

\subsection{Coefficient of thermal sensitivity}

\begin{figure*}[tbh]
	\begin{centering}
		\includegraphics[width=0.48\textwidth]{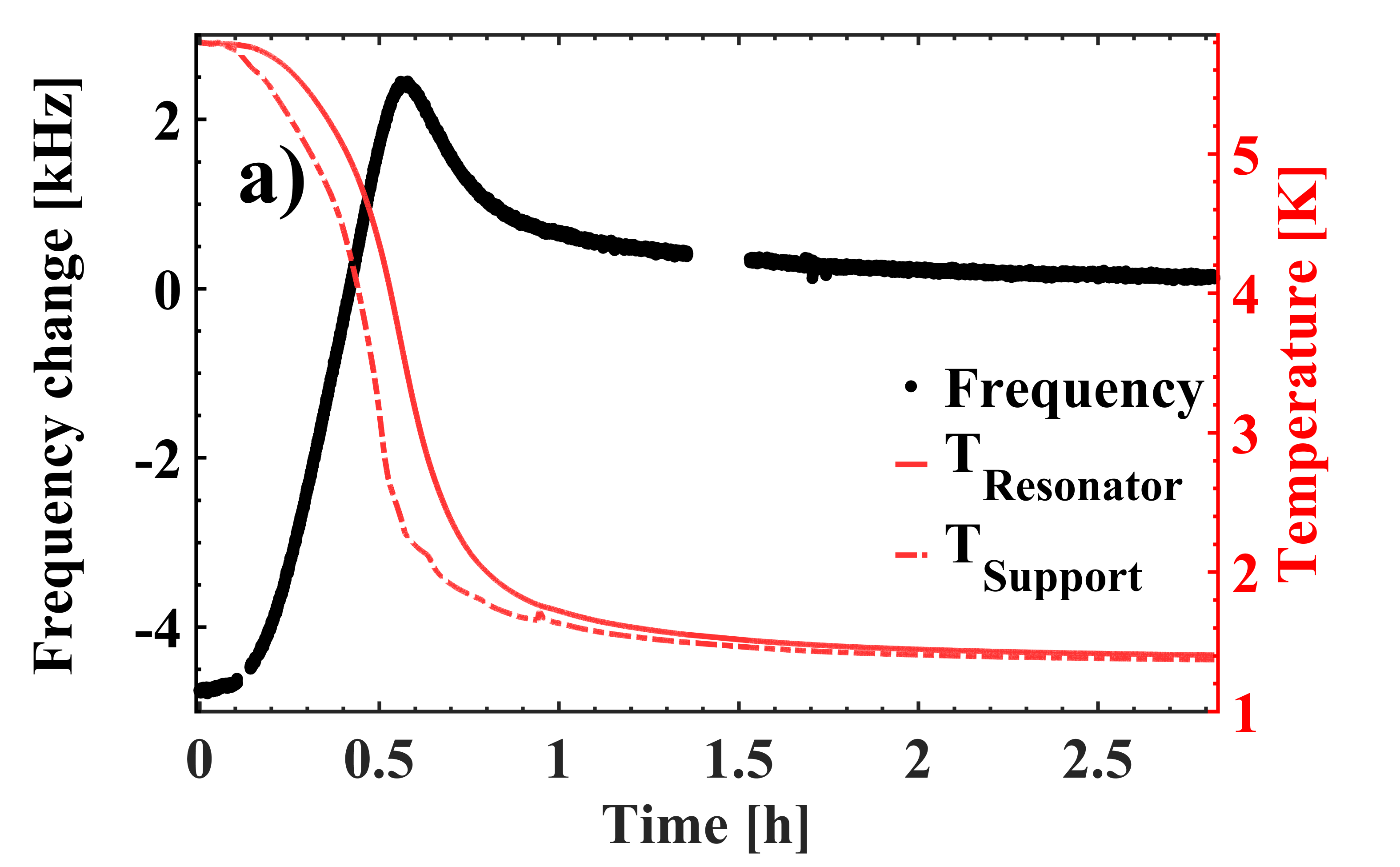}
		\includegraphics[width=0.48\textwidth]{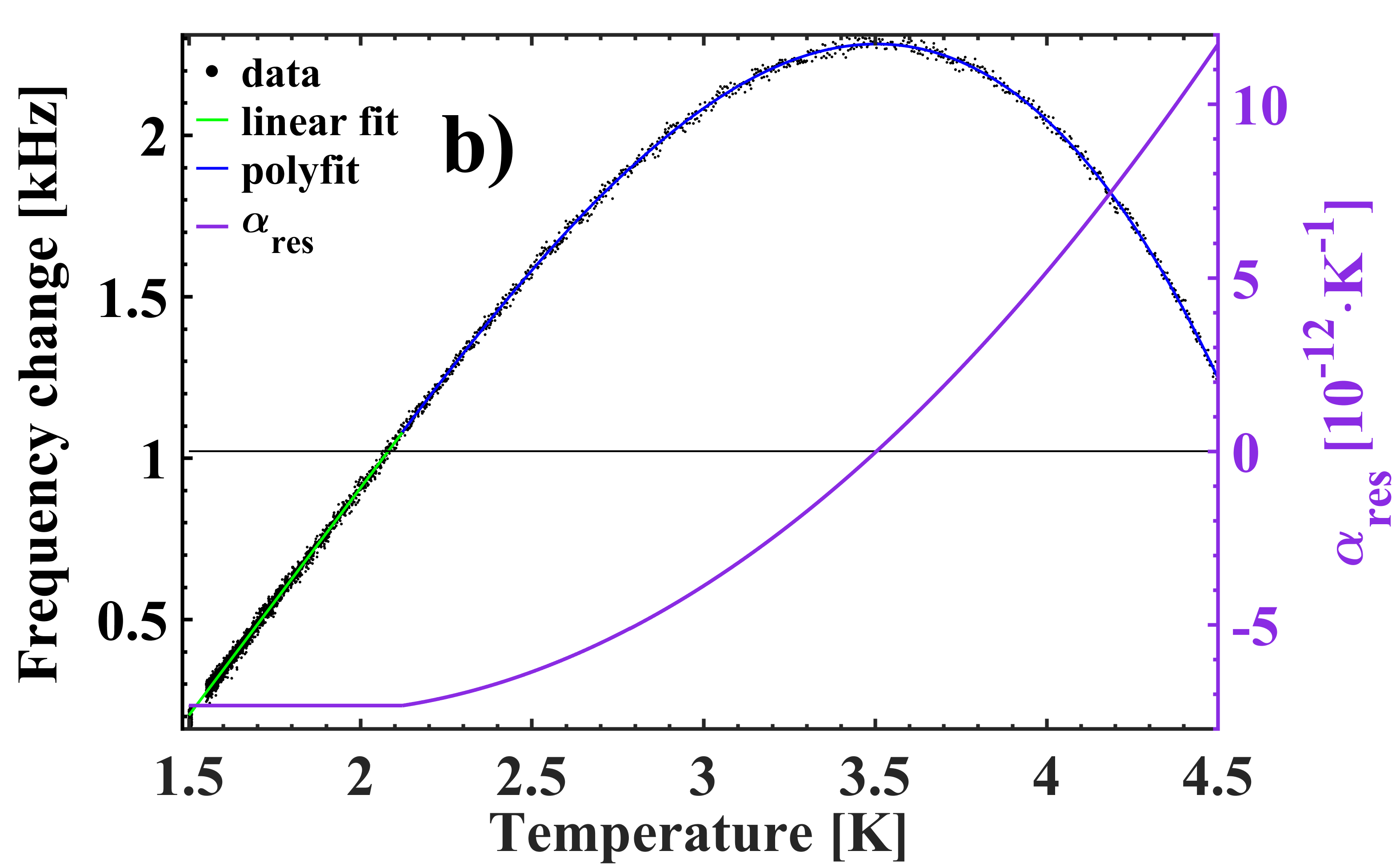}
		\includegraphics[width=0.48\textwidth]{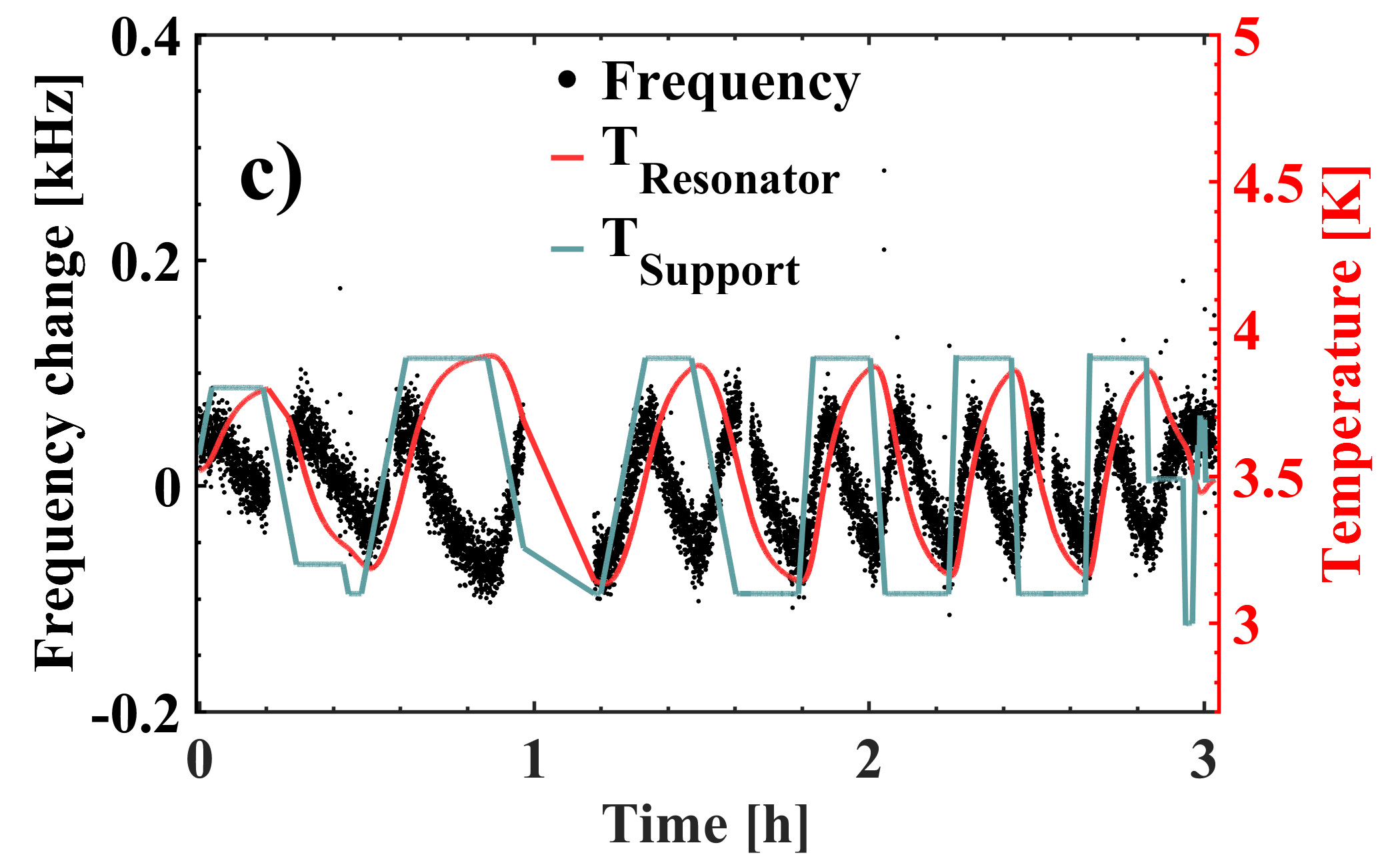}
		\includegraphics[width=0.48\textwidth]{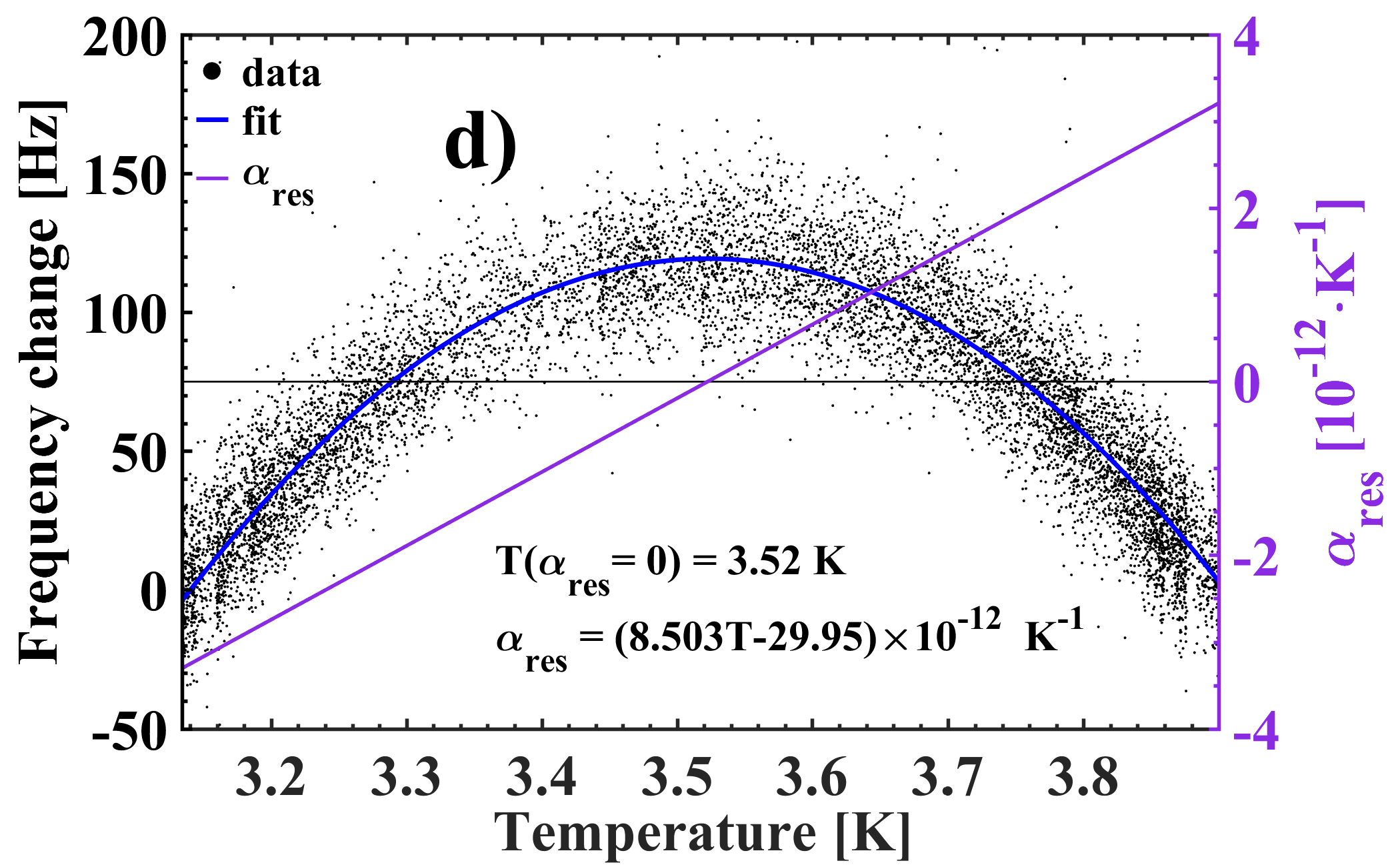}
		\includegraphics[width=0.48\textwidth]{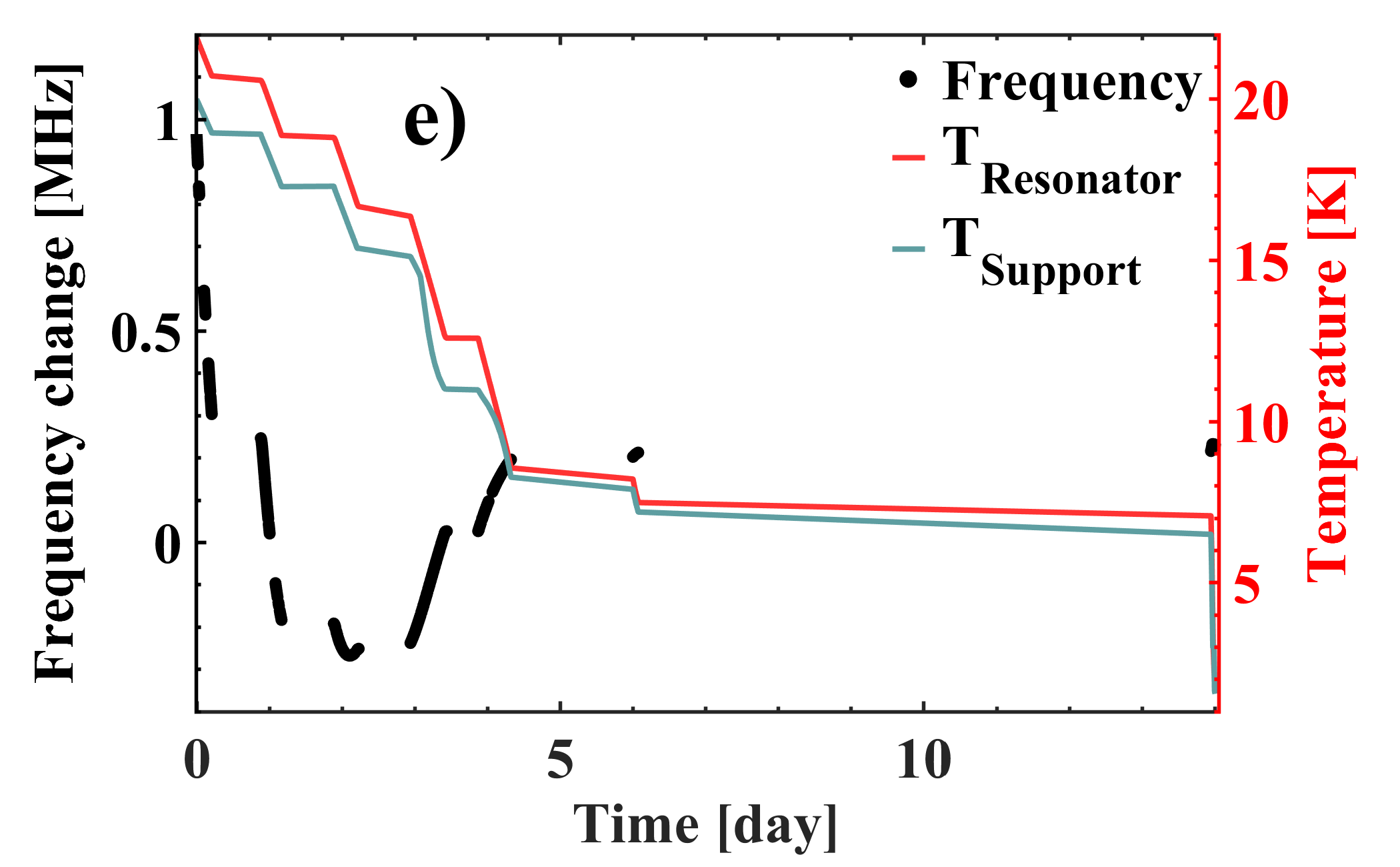}
		\includegraphics[width=0.48\textwidth]{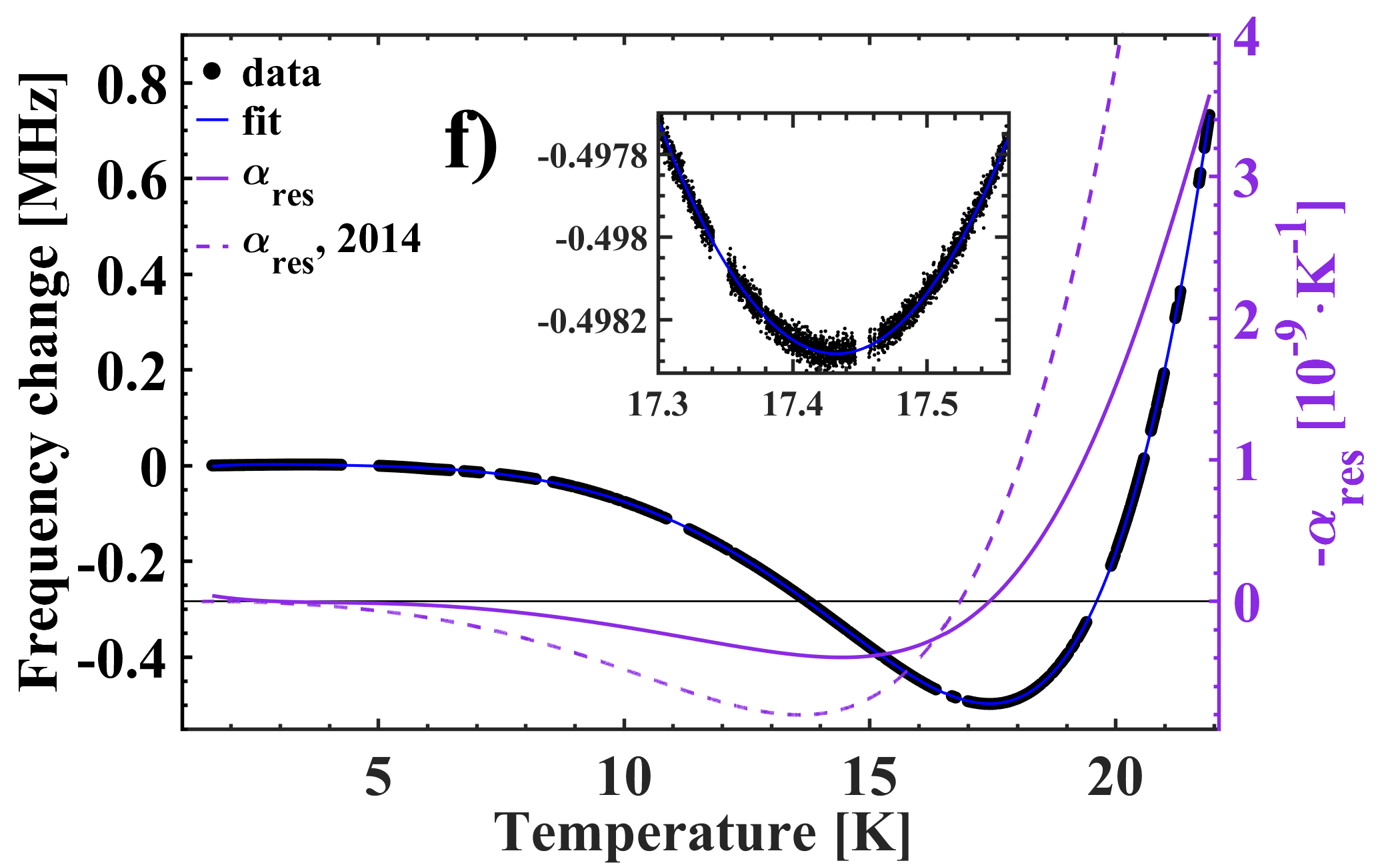}
		\par\end{centering}
\caption{\label{fig:Si5-CTE}Measurements of the thermally induced frequency
change of the resonator in three temperature ranges: (a,~b) between
$1.5$~K to $4.5$~K; (c,~d) around $T_{0}=3.52$~K (zero CTF);
(e,~f) from $22$~K to $1.5$~K. The red dashed line in (f) is
the result of Ref.~\onlinecite{Wiens2014}.}
\end{figure*}

The coefficient of thermal sensitivity of frequency (CTF) of our resonator,
$\alpha_{\mathrm{res}}$ was determined in the temperature range between
$5.5$~K to $1.5$~K by cooling down or heating up the cryostat
over several hours and simultaneously measuring the resonator frequency
using the half-transmission detuning technique described above. The
raw data of the experiments are presented in Fig.~\ref{fig:Si5-CTE}~a,
c, and e. The total change in resonator frequency from 5.5~K to 1.5~K
is $7$~kHz. Clearly, the resonator frequency exhibits and extremum
at 3.52~K, its CTF vanishes there.

The CTF in the temperature region below $2.12$~K was determined
by a linear-in-$T$ fit to the data (see Fig.~\ref{fig:Si5-CTE}~b).
The resulting CTF is constant, $\alpha_{\mathrm{res}}(T<2.1\,{\rm K})=-7.33\times10^{-12}{\rm K}^{-1}$
with a fit error smaller than $0.3$\%. The remaining data were fitted
with a cubic polynomial (see Fig.~\ref{fig:Si5-CTE} b). We find
$\alpha_{\mathrm{res}}(2.1\,{\rm K}<T<4.5\,{\rm K})=(2.72(T/{\rm K})^{2}-10.0\,(T/{\rm K})+1.65)\times10^{-12}{\rm K}^{-1}$.
While the uncertainty of the $T^{2}$ coefficient is 2\%, it is smaller
than $1$\% for the two remaining coefficients.

In order to determine precisely the temperature of zero CTF, a sinusoidal
modulation of the temperature was applied around the mean temperature
of $T=3.5$~K, using a heater attached to the experimental plate.
The corresponding change in frequency of the resonator is depicted
in Fig.~\ref{fig:Si5-CTE}~c. A quadratic fit was performed (see
Fig.~\ref{fig:Si5-CTE}~d), yielding a zero CTF temperature $T_{0}=3.52\pm0.02$~K
with derivative $d\alpha_{\mathrm{res}}/dT=8.5\times10^{-12}/\mathrm{K}^{2}$.
This value is a factor $40$ smaller than at the zero-CTF-temperature
$17.4$~K, where it is $-3.4\times10^{-10}/\mathrm{K}^{2}$ (see
below), and a factor $2000$ lower than at $124.2$\,K, $1.7\times10^{-8}/\mathrm{K}^{2}$,
(Ref.~\onlinecite{Kessler2012a}). Assuming that the operating temperature has
an undesired offset of 0.02~K from the zero-CTF temperature, the
CTF is a factor 35 smaller than the CTF at $1.5$~K.

We also determined the CTF in the extended temperature range between
$22$~K and $T_{0}$ from data obtained during a $15$-day-long slow
cool-down of the cryostat. The results are presented in Fig.~\ref{fig:Si5-CTE}~e,~f.

Our results on the CTF for the temperature range below 6~K were confirmed
on different occasions: upon heating and cooling of the setup, after
prolonged operation ($>0.5$ year) at $1.5$~K and twice immediately
after heating of the whole setup to over $100$~K. During these measurements
we found a minor variation of the zero CTF temperatures and of the
slope at this temperature: $\pm20$~mK and $\pm1\times10^{-12}/\mathrm{K}^{2}$,
respectively.

The CTF results differ significantly from the data published in Ref.~\onlinecite{Wiens2014},
which was obtained for a silicon resonator with a different support.
While in that work the supporting frame was also made of copper, the
resonator was held by ten flexible steel wires that reduced strongly
the influence of the thermal expansion of the copper frame. In the
present work, the resonator and the copper frame are connected by
friction. Thus, a temperature change of the whole setup may conceivably
cause a strain $\epsilon=\Delta R/R$ along the radial direction of
the resonator, due to the much higher expansion coefficient of the
copper support, $\alpha_{\mathrm{Cu}}(T=3.52\,{\rm K})=2.3\times10^{-9}/{\rm K}$
(Ref.~\onlinecite{McLean1969}) compared to $\alpha_{\mathrm{Si}}(T=3.52\,{\rm K})=2.1\times10^{-11}/{\rm K}$ (Ref.~\onlinecite{Wiens2014}).
This strain is converted into a change of the distance between the
mirrors $\Delta L/L$ via the material's Poisson's ratio $(\Delta L/L)=-\epsilon\,\nu$.
This hypothesis can be tested by considering the CTF at higher temperatures,
where the difference between the CTF of silicon and of copper is larger.
For example, at $20$~K $\alpha_{\mathrm{Si}}(T=20\,{\rm K})=4\times10^{-9}/{\rm K}$
(Ref.~\onlinecite{Wiens2014}) and $\alpha_{\mathrm{Cu}}(T=20\,{\rm K})=0.27\times10^{-6}/{\rm K}$ (Ref.~\onlinecite{ReedClark1983}). However, we find a CTF similar to our previous work (see Fig.~\ref{fig:Si5-CTE}~f).
The second zero-CTF temperature is $T_{\text{0}}'=17.4\,{\rm K}$,
compared to our earlier $T_{0}'=16.8\,{\rm K}$.

Thus, more detailed studies are necessary to determine the precise
reason for the zero crossing, including FEM simulations and measurements
with different implementations of the contact surfaces between frame
and resonator.

\section{Frequency stability}

We measured the stability of the laser frequency when referenced to
the silicon resonator using both techniques outlined above.

\subsection{Medium-term frequency instability}

\begin{figure*}[tbh]
	\begin{centering}
		\includegraphics[width=0.48\textwidth]{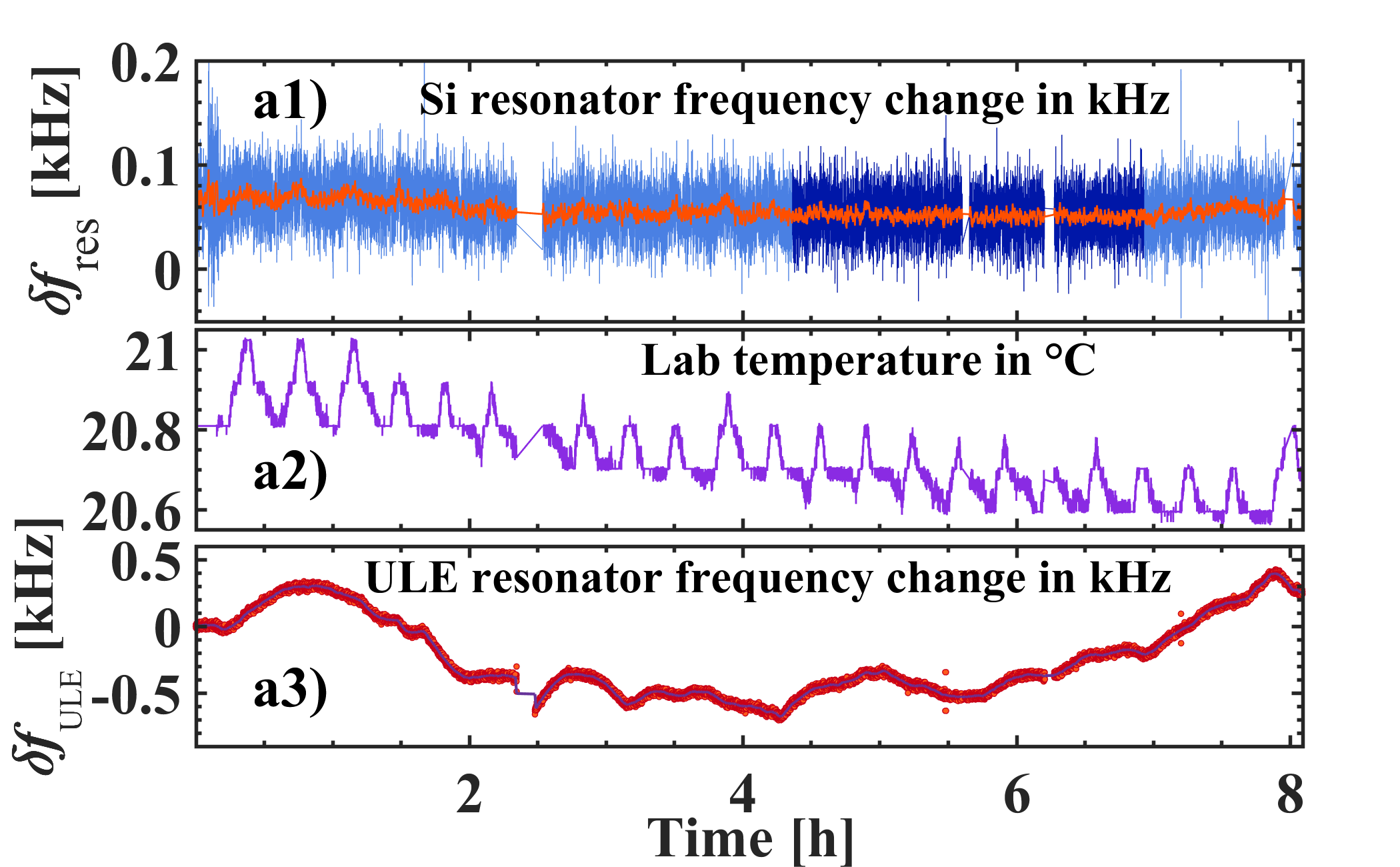}
		\includegraphics[width=0.48\textwidth]{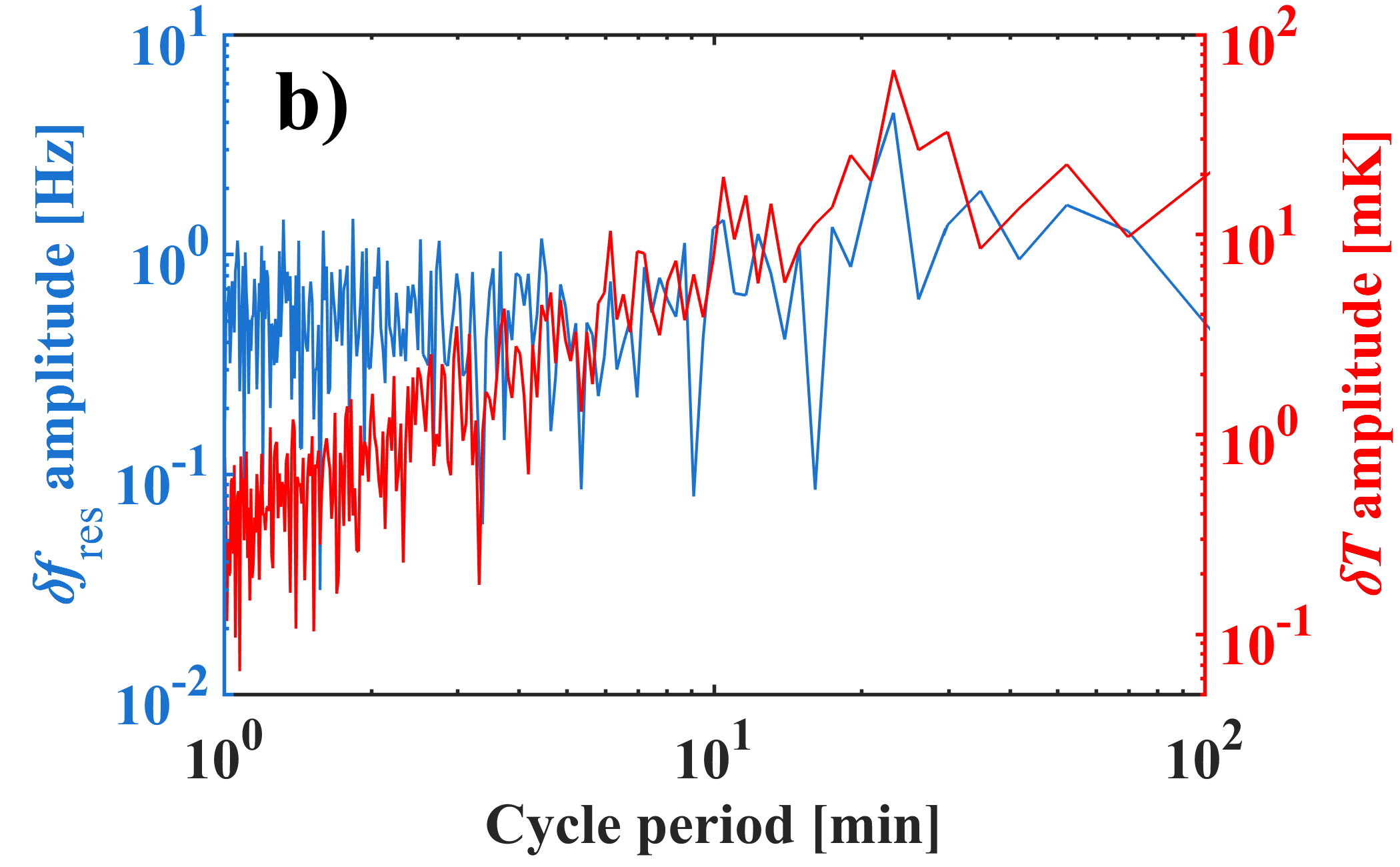}
		\includegraphics[width=0.48\textwidth]{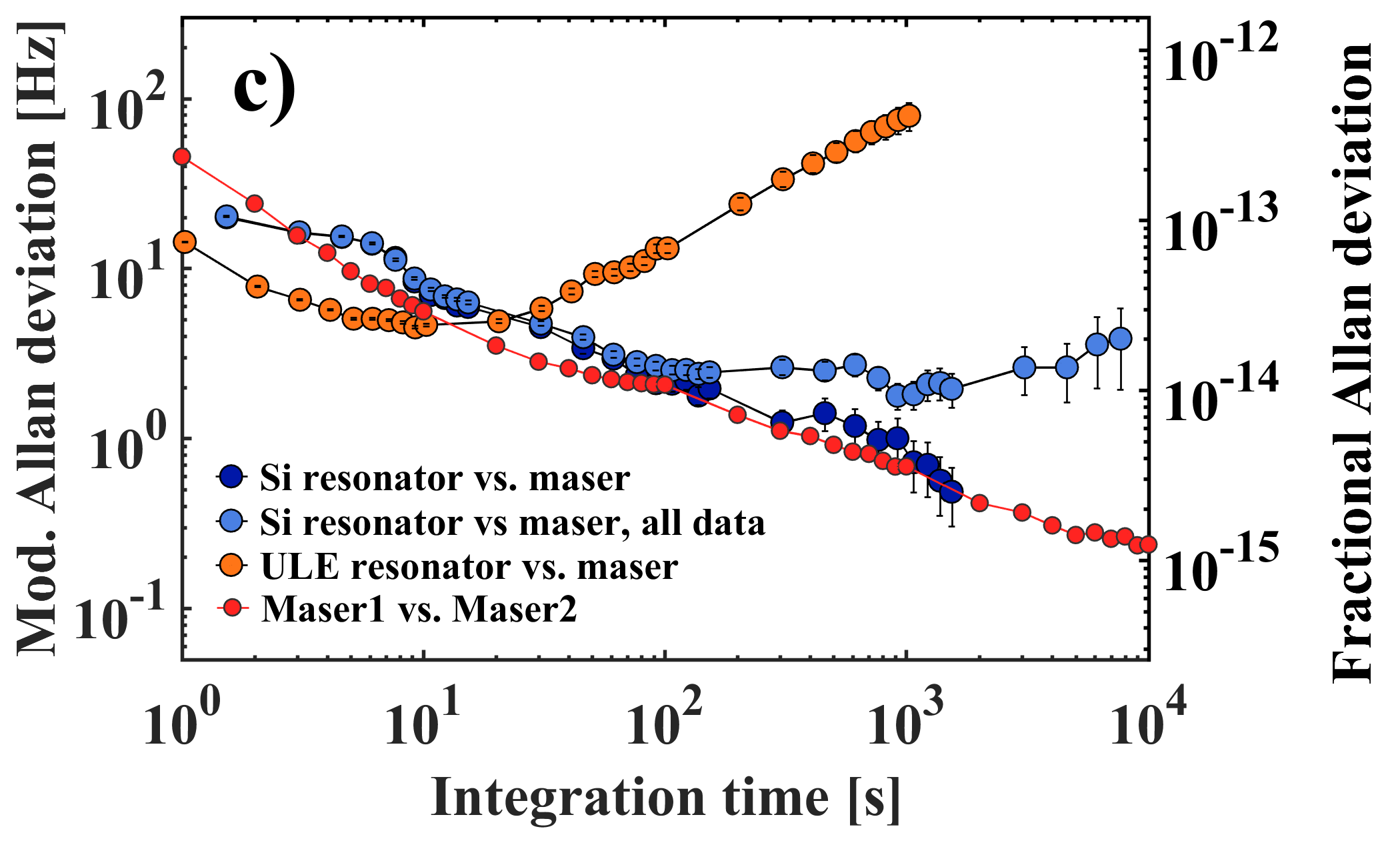}
		\par\end{centering}
	\caption{\label{fig:Si5-Frequency-Stability} (a) An eight-hour-long laser
		frequency measurement, with respect to the maser, when the frequency-shifted
		laser wave tracks the silicon resonator (plot a1). The dark colored
		interval is used for the calculation of Allan deviation in panel (c).
		Plots a2 and a3 show the temperature of the lab housing the cryostat
		and the frequency change of the pre-stabilized laser used for interrogation
		of the silicon resonator, respectively. (b) FFT of the lab temperature
		and resonator frequency time traces presented in panel (a). (c) Modified
		Allan deviation of the silicon resonator frequency (whole dataset
		(light blue markers) and selected part (dark colored blue markers))
		and of the pre-stabilized laser (orange markers). No drift was removed.
		The red points in (c) show the maser instability, determined from
		a measurement of the frequency difference of two nominally identical
		masers on a different occasion.}
\end{figure*}

The result of a frequency measurement, allowing to determine
the medium-term frequency instability, and obtained with the linescan
technique, is presented in Fig.~\ref{fig:Si5-Frequency-Stability}~a.
Here, the temperature of the resonator was at $1.47$~K and was
not actively stabilized. As seen in Fig.~\ref{fig:Si5-Frequency-Stability}~a,
the frequency stability of the resonator is compromised by the periodic
variations of the laboratory temperature. This can be more clearly
seen in the FFT spectrum of frequency and temperature time traces,
presented in Fig.~\ref{fig:Si5-Frequency-Stability}~b. We observe
a modulation of frequency with an amplitude of $4$~Hz at a time
period of $23$~min, which is identical to the duration of the laboratory
temperature variations. Therefore, we only consider the most stable
part of the data, exhibiting the lowest drift, and computed the modified
Allan deviation (see Fig.~\ref{fig:Si5-Frequency-Stability}~c).
We find that for integration times up to $1000$~s the laser frequency
instability is slightly higher than the beat instability of our two active
hydrogen masers and approaches the minimum value of $0.5$~Hz ($2.5\times10^{-15}$)
at $\tau=1500$~s. It is possible that this level has a contribution
from the temperature variations of the resonator, see Fig.~\ref{fig:Temperature-Instability}~a.

\begin{figure*}[tbh]
	\begin{centering}
		\includegraphics[width=0.48\textwidth]{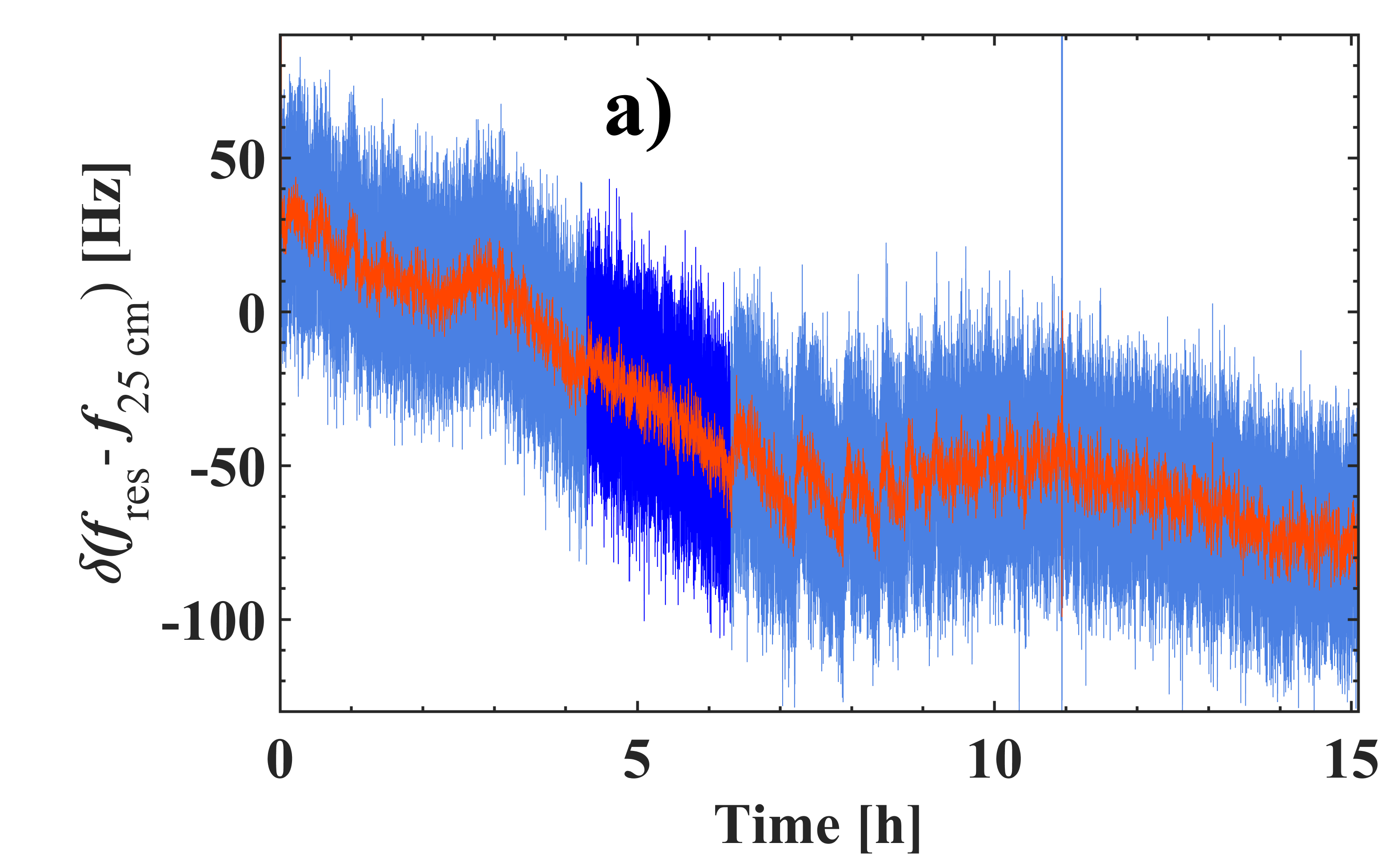}
		\includegraphics[width=0.48\textwidth]{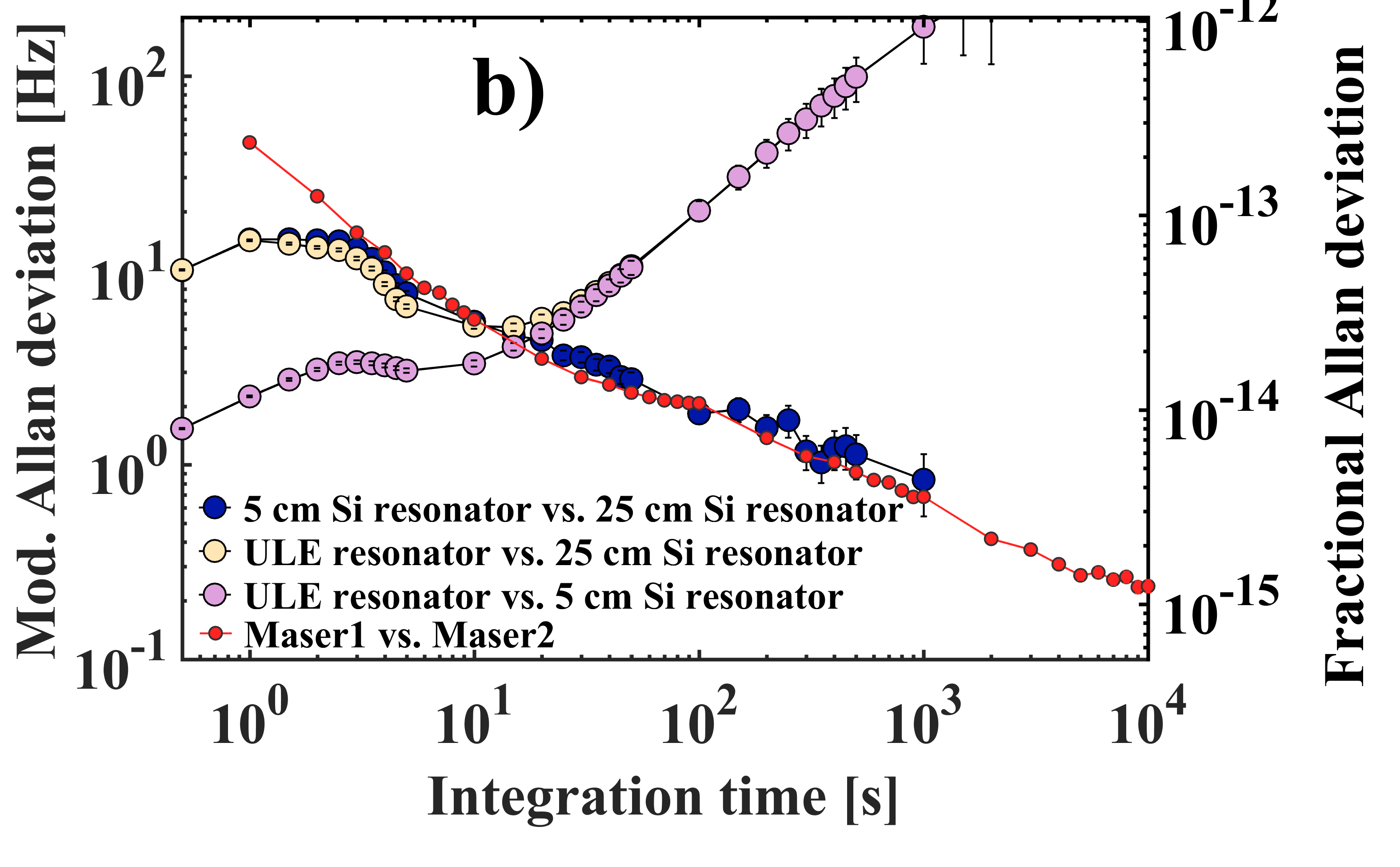}
		\par\end{centering}
	\caption{\label{fig:Si5-Frequency-Stability-From-Beat-with-Si1} (a) A fifteen-hour-long measurement of the frequency
		of the $5$~cm silicon resonator relative to the frequency of the $25$~cm silicon resonator. The two-hour-long dark colored interval is used for the calculation of Allan deviation. (b) Modified Allan deviation of the frequency difference (selected part in panel (a)) between the silicon resonator and the $25$~cm silicon resonator (dark blue markers)  A drift of $5$~mHz/s was removed. Purple and yellow markers represent the frequency instability of the silicon resonator and the $25$~cm silicon resonator relative to the ULE resonator, respectively, measured simultaneously. Red points show the maser instability, determined from a measurement of the frequency difference of two nominally identical masers 1 and 2 on a different occasion.}
\end{figure*}

To verify our results presented in Fig.~\ref{fig:Si5-Frequency-Stability}~c we apply the half-amplitude technique to stabilize the ULE laser light to both the $5$~cm resonator and the $25$~cm silicon resonator, using two independent AOMs (see schematic in Fig.~\ref{fig:Schematic-Optical-Setup}). We use the frequency difference between these two resonators for the estimation of frequency instability. This procedure allows us to eliminate fluctuations of the ULE frequency and to avoid an introduction of potentially present additional noise coming from the frequency comb. The result of the measurement is presented in Fig.~\ref{fig:Si5-Frequency-Stability-From-Beat-with-Si1}. To determine the frequency instability of the $5$~cm resonator we use the most stable part with a duration of $2$~h. The calculated modified Allan deviation of this part is presented in Fig.~\ref{fig:Si5-Frequency-Stability-From-Beat-with-Si1} b.  

For integration times  $\tau\leqslant10$~s the resulting instability is below the instability of the maser. The beat with the ULE resonator shows an instability of $2$~Hz ($1\times10^{-14}$) at $1$~s integration time. 

For integration times from $10$~s to $1000$~s the instability of the Si5-Si1 beat follows closely the instability of the maser-maser beat and approaches the minimum instability of $0.8$~Hz or $4\times10^{-15}$ at $\tau=1000$~s integration time. With the assumption that both resonators contribute equally to the instability we can divide the above result by a factor $\sqrt{2}$ and obtain $0.6$~Hz ($3\times10^{-15}$) at $\tau=1000$~s integration time.     

\section{Long-term frequency drift}

The long-term ($\tau>10~000$~s) frequency drift of the resonator is mainly determined
by the length changes of the spacer due to the relaxation processes
in the crystal lattice, and photochemical changes in the mirror coatings.
The latter are found to depend on the cumulative irradiation duration
by the laser and the applied laser power. In our previous work with
horizontally oriented silicon resonators supported as described above,
we observed an exponential relaxation of the frequency and a positive
drift rate after a long time \cite{Wiens2016}. In those measurements,
we interrogated the resonator with a pre-stabilized laser wave of
$1$~$\mu$W power for a duration of $1$~h each day and blocked the
laser light for the remaining $23$~h.

\begin{figure*}[tbh]
	\begin{centering}
		\includegraphics[width=0.78\textwidth]{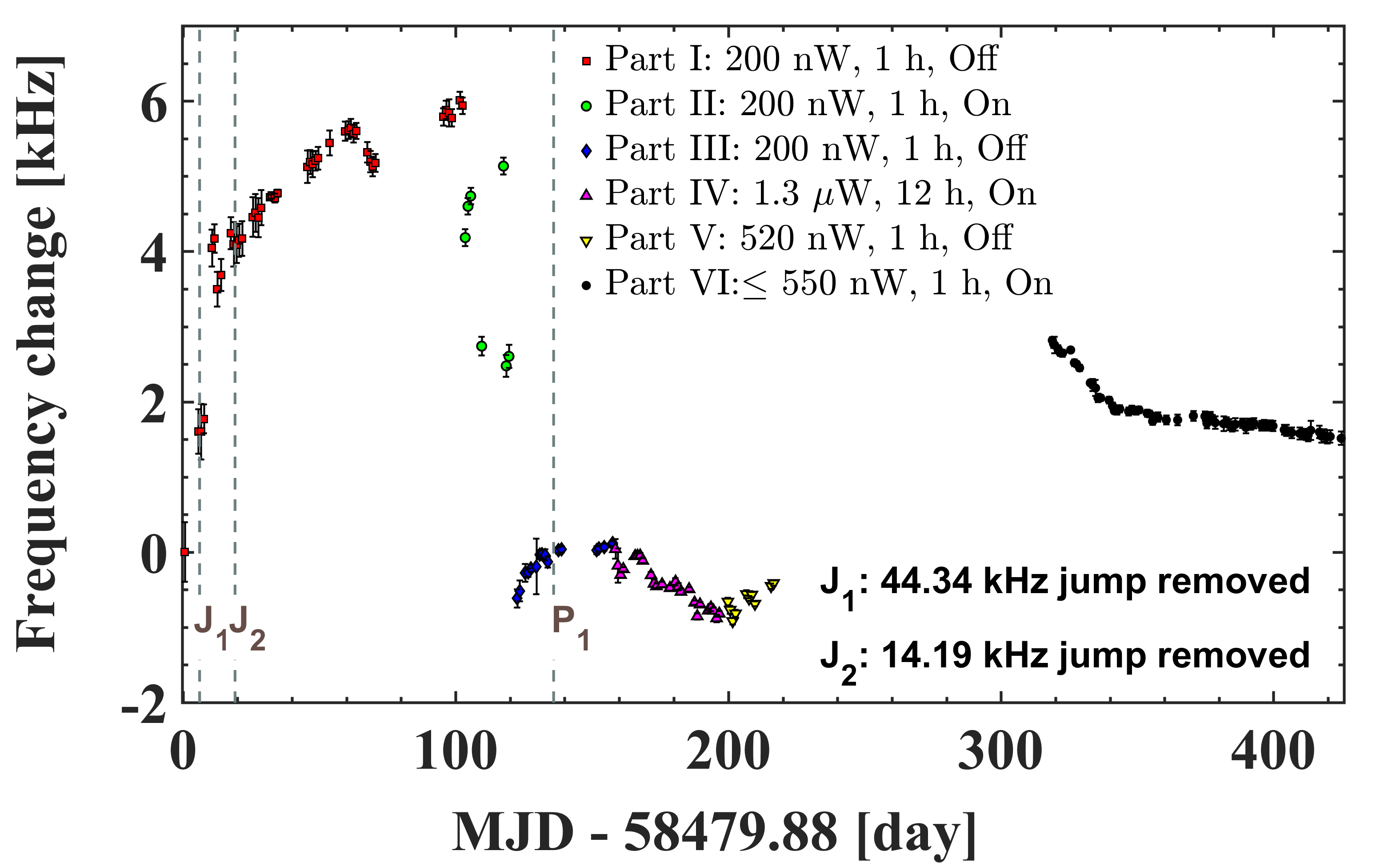}
		\par\end{centering}
\caption{\label{fig:Si5-Long-Term-Frequency-Drift-All}Frequency change of the resonator measured over a time span of $425$~days. Between day $217$ and $318$ no data was recorded, because the temperature of the resonator was changed to measure the coefficient of thermal sensitivity. $\mathbf{\mathrm{J_{1}}}$ and $\mathbf{\mathrm{J_{2}}}$ mark two frequency jumps that occurred during the measurement, $\mathrm{P}_{1}$ marks the time when we changed from the line scan technique to the half-amplitude detuning technique. The legend sates, for each part: the incident optical power of the laser during the measurements, the measurement duration per working day, and whether the laser light was on or off in-between measurements on subsequent days.}
\end{figure*}

The long-term frequency drift rate of the $5$~cm resonator was determined
by repeatedly scanning over its resonance line with the pre-stabilized
laser light, over a time period of $425$~days, starting at a day
zero (MJD 58479.88), when the system reached the temperature of $1.5$~K. The results
are presented in Fig.~\ref{fig:Si5-Long-Term-Frequency-Drift-All}.
We subtracted two frequency jumps, marked as $\mathbf{\mathrm{J_{1}}}$
and $\mathbf{\mathrm{J_{2}}}$. They were probably caused by spontaneous
relaxation processes in the crystalline spacer and/or in the substrates.

In order to gain insight into the drift behaviour, we varied the incident laser power. It was kept at $200$~nW for all measurements up to day $154$. From the following day and up to day $197$ the optical power was increased to the level of $1.3$~$\mu$W. It was reduced again to $520$~nW for the days up to day $216$. From day $320$ to day $425$ the power was subsequently reduced from $550$~nW to $100$~nW. On five days between day $102$ to day $122$ we optimized the incoupling of the laser light to the lower mirror of the resonator for the installation of the Pound-Drever-Hall setup attached to the vacuum enclosure of the cryostat. The incident light power on these occasions was on the order of $1$~mW.

To determine the frequency of the resonator we used the linescan technique from day $1$ to day $137$. After this day, marked as $P_{1}$ in the diagram, we changed to the half-transmission technique. The daily interrogation duration was approx. $1$~h on all working days except for two time periods from day $159$ to day $197$ and from day $320$ to day $425$ when it was increased to $12$~h and to $8$~h on average, respectively. The laser light was blocked between the measurements from day $1$ to day $158$. For the two mentioned time periods from day $159$ to day $197$ and from day $320$ to day $425$ the laser light was not at resonance when no measurements were performed (i.e. for 12~h per working day, 24~h on the week-end days) but it was still incident on the front mirror of the resonator.

\begin{figure*}[tbh]
	\begin{centering}
		\includegraphics[width=0.48\textwidth]{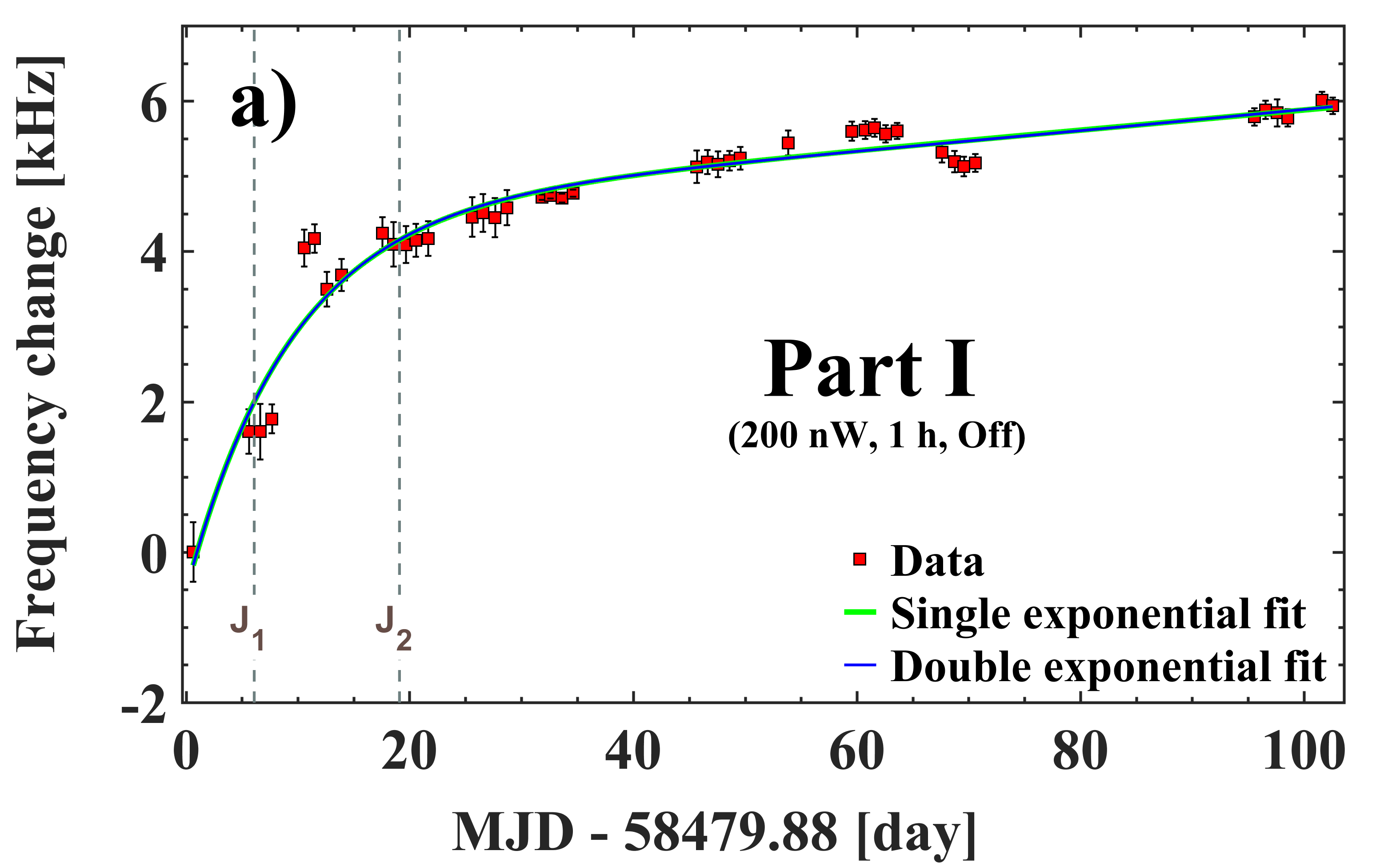}
		\includegraphics[width=0.48\textwidth]{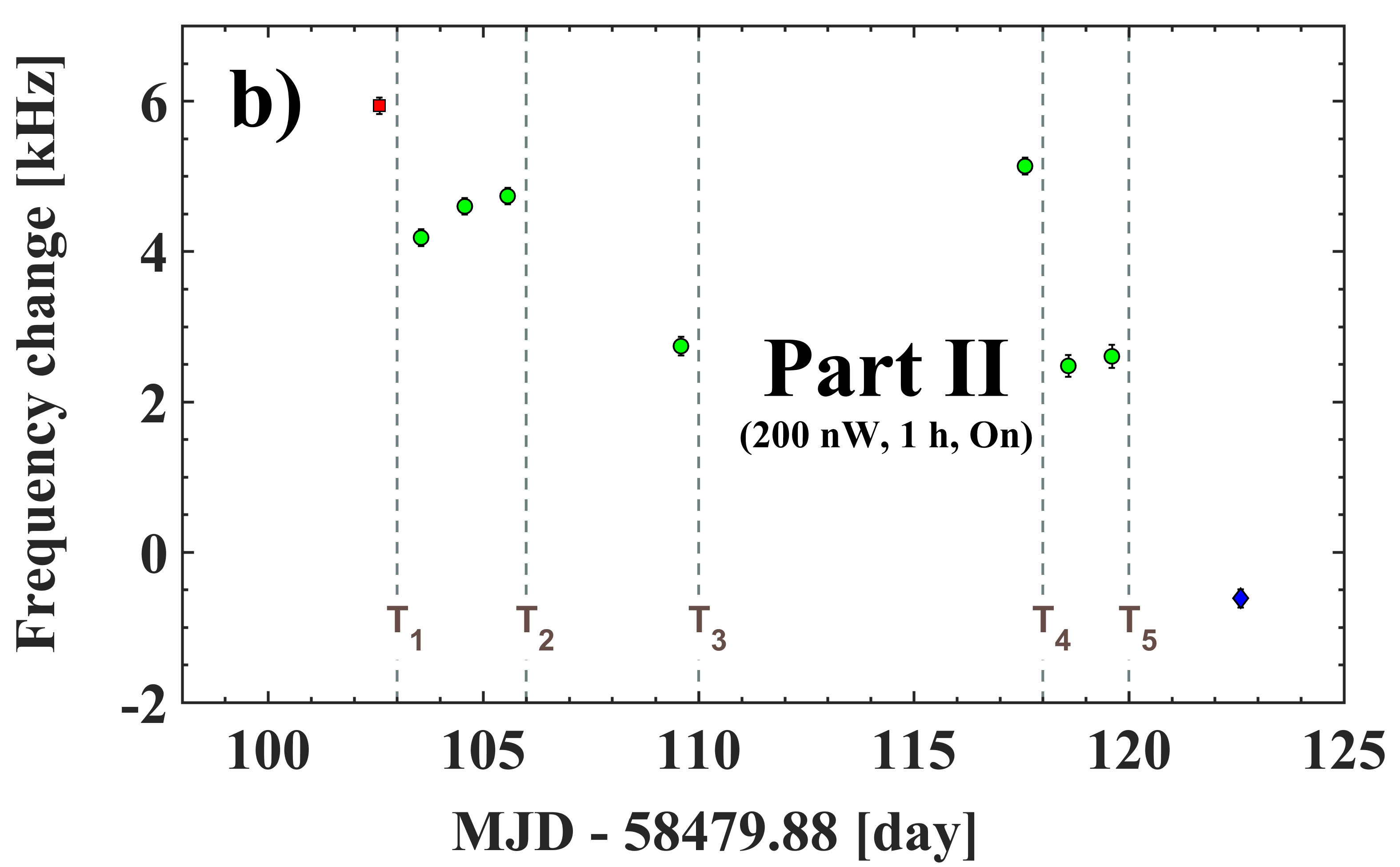}
		\includegraphics[width=0.48\textwidth]{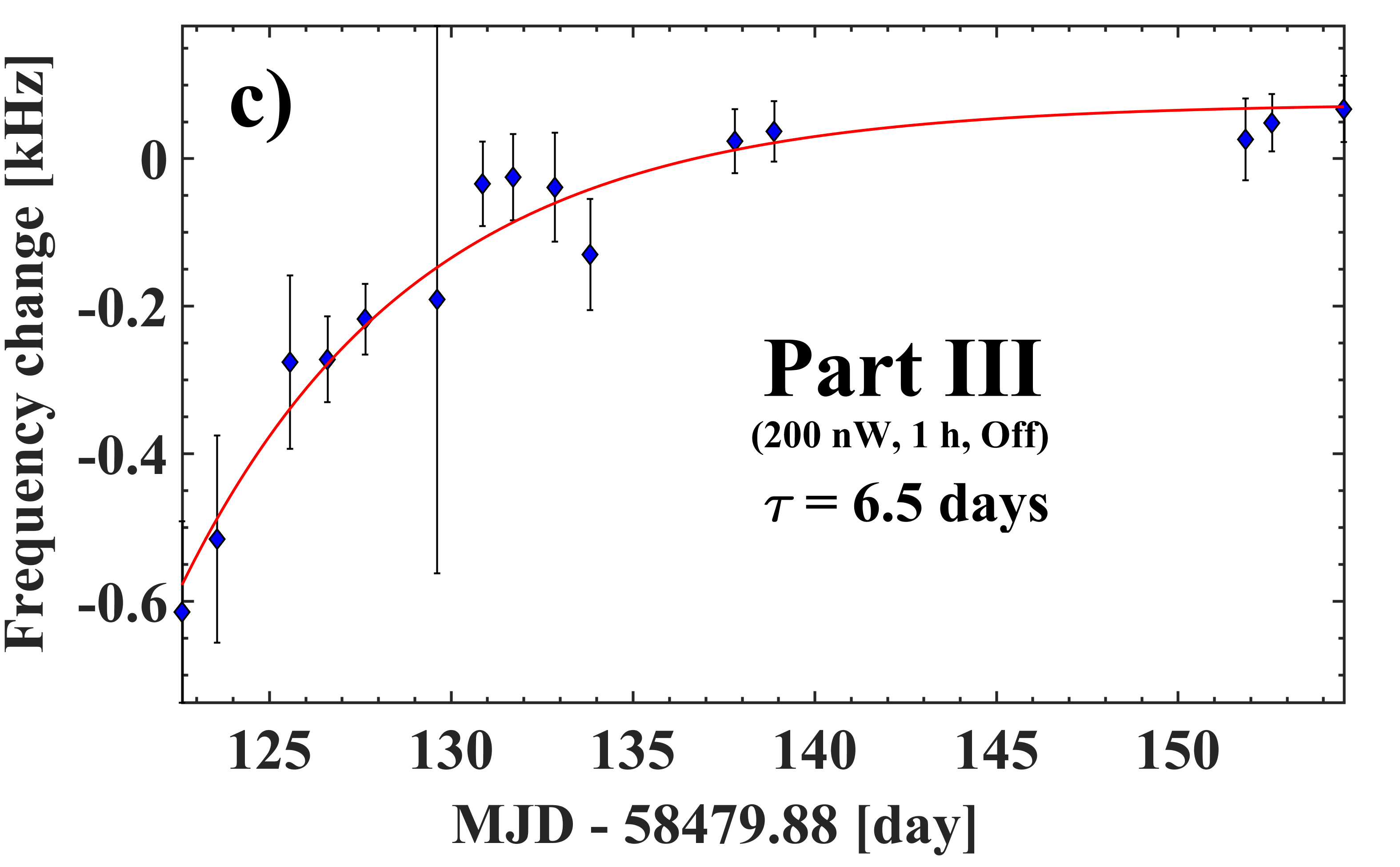}
		\includegraphics[width=0.48\textwidth]{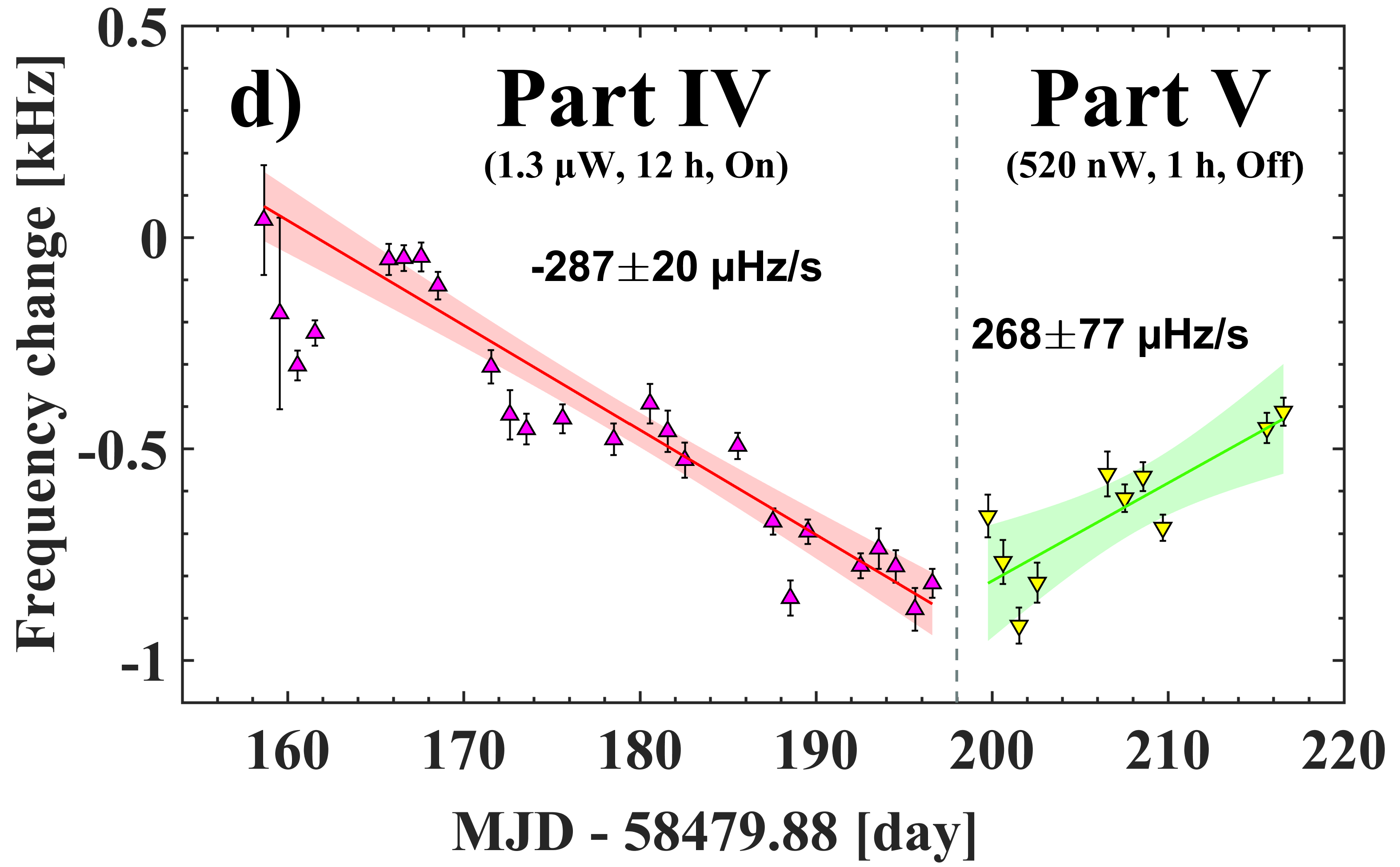}
	\par\end{centering}
\caption{\label{fig:Si5-Long-Term-Frequency-Drift} Zooms into different intervals
	of the long-term frequency measurement campaign: (a) Part I; (b) Part
	II: $\mathbf{\mathrm{T_{1}}}$ to $\mathbf{\mathrm{T_{5}}}$ mark
	the times when a separate 1~mW free-space laser beam was coupled
	into the resonator on the bottom side. In addition, regular 1~hour-long
	daily frequency measurements were performed with an incident optical
	power of $200$~nW; (c) Part III: subsequent relaxation (interrogation
	at low power); (d) Part IV and Part V. Linear fits and corresponding
	drift rates are shown.}
\end{figure*}

Starting immediatly after cool-down, we observed a relaxation of the
resonator frequency with a measured total frequency change of 6~kHz
over the first $100$ days. We denote this measurement period as Part
I in Fig.~\ref{fig:Si5-Long-Term-Frequency-Drift}. The frequency
change is positive, meaning that the distance between the two mirrors
is decreasing with time. The rate of frequency change is not constant
over time.

At the end of Part I, the drift rate was $9.7\times10^{-19}$~/s.
This value is a factor three higher then the drift measured on a resonator
with comparable dimensions by Ref.~\onlinecite{Robinson2019} and a factor seventy
higher compared to our previous result obtained with a $25$~cm long
silicon resonator \cite{Wiens2016}. However, this latter drift was
measured after almost a year-long continuous operation at $1.5$~K.
The drift after the first $100$ days was almost identical to the current
result, $1\times10^{-18}/$s.

During Part II (Fig.~\ref{fig:Si5-Long-Term-Frequency-Drift-All}~b),
on five occasions with a duration of approx. $4$~h each (marked as $\mathbf{\mathrm{T_{i}}}$
in the figure), a laser beam with $1$~mW optical power was
coupled into the resonator mode for most of that time, through a cryostat
window and into the bottom end of the resonator. Frequency measurements
during this time interval (performed as usual) display negative frequency
changes. After each optimization of the PDH setup there is a frequency
jump on the order of $-2$~kHz to $-3$~kHz. These jumps have different
sign, compared to two jumps $\mathbf{\mathrm{J_{1}}}$ and $\mathbf{\mathrm{J_{2}}}$
(see Fig.~\ref{fig:Si5-Long-Term-Frequency-Drift-All}~a), suggesting
that the two underlying processes are different in nature.

After completion of the optimization we observe an exponential relaxation
process with $\tau=6.6$~days over the next month (Part III, see
Fig.~\ref{fig:Si5-Long-Term-Frequency-Drift}~c). As can be seen
in Fig.~\ref{fig:Si5-Long-Term-Frequency-Drift-All}, the resonator
frequency did not return to the original value, implying that the
changes induced by high-power irradiation were permanent.

Fig.~\ref{fig:Si5-Long-Term-Frequency-Drift}~d displays the next
two parts of the measurement campaign, Part IV and Part V. The frequency
drift rate changes from negative to positive after reduction of power
from $1.3$~$\mu$W to $520$~nW and interrogation time from $12$~h
to $1$~h.
\begin{figure}[tbh]
	\begin{centering}	
			\includegraphics[width=0.48\textwidth]{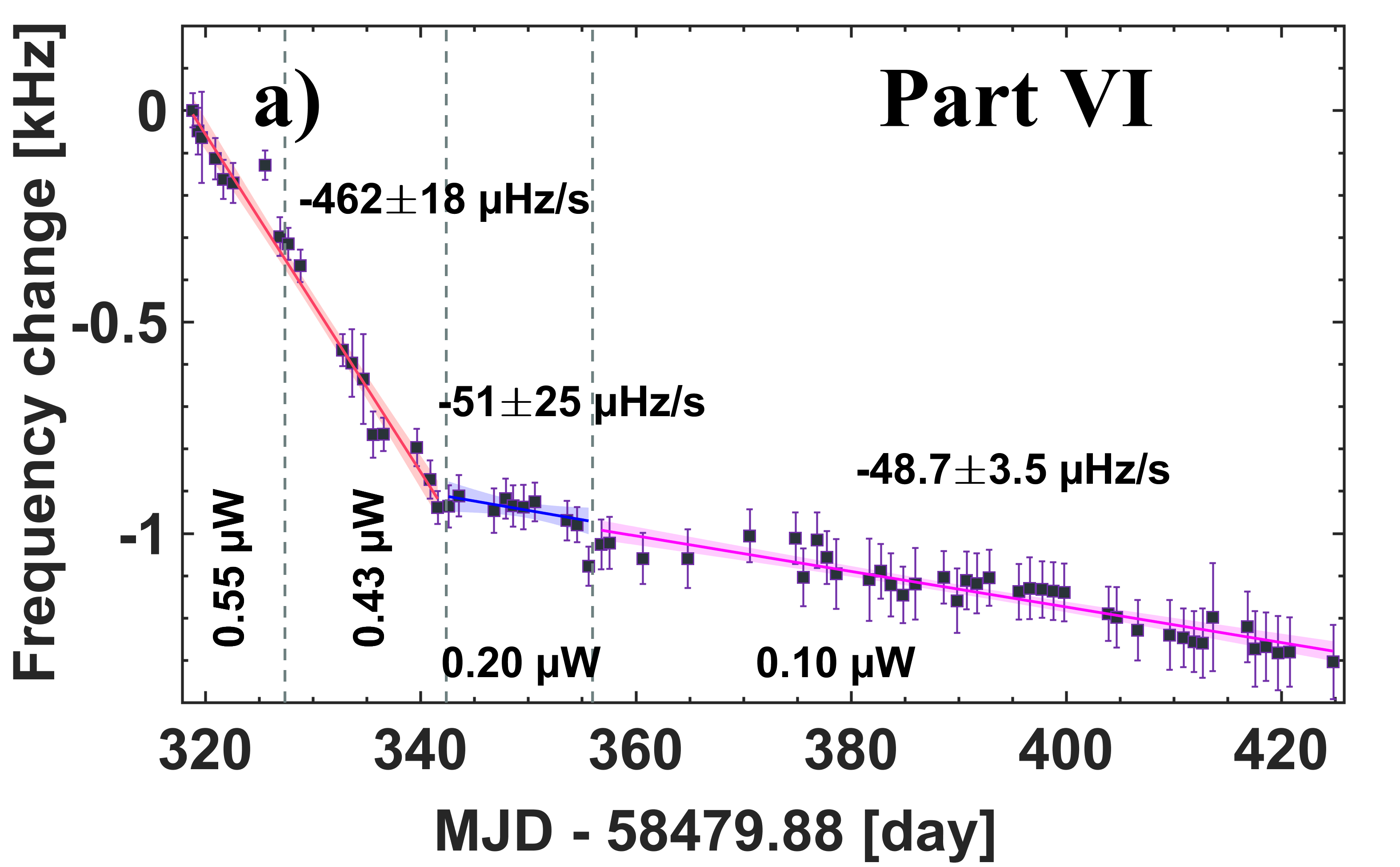}
			\includegraphics[width=0.48\textwidth]{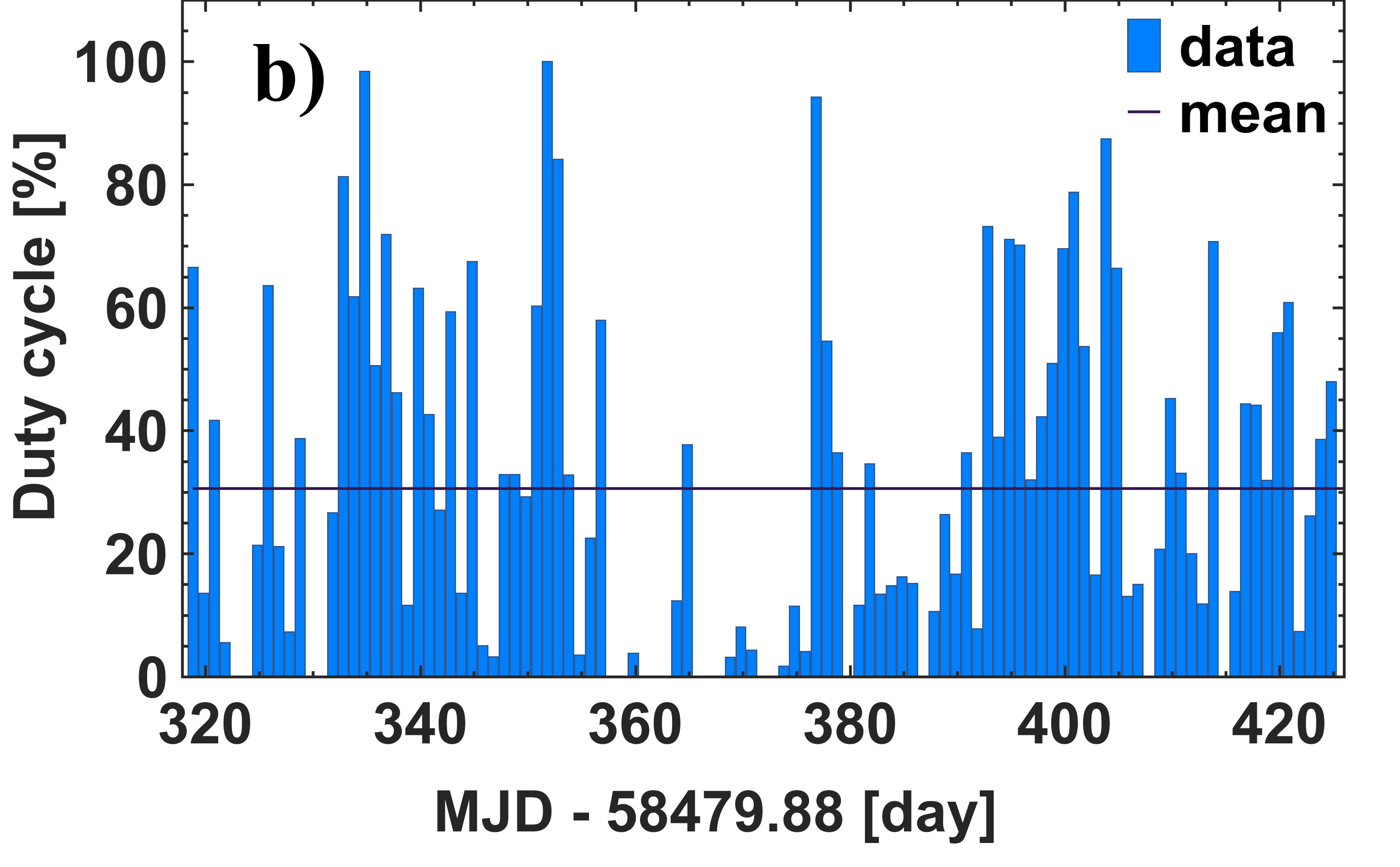}
	\par\end{centering}
	\caption{\label{fig:Frequency-Drift-On-Power} (a) Frequency change of the resonator measured over a time span of $105$~days as a function of optical power. (b) The duty cycle of the resonator interrogation in percent per day.}
\end{figure}
During Part VI we performed a systematic study of the frequency drift dependence on the optical laser power, where we continuously decreased the level of optical power from $550$~nW to $100$~nW (see Fig.~\ref{fig:Frequency-Drift-On-Power} a). To simulate the realistic conditions of a day-to-day operation of the resonator we increased the duty cycle to an average duration of $8$~h per day (see Fig.~\ref{fig:Frequency-Drift-On-Power} b). To exclude possible thermal influences we actively stabilized the temperature at $1.4$~K during this time period. In line with the measurements in Part VI and Part V we observe a decrease of the frequency drift with reduction of optical power. The lowest drift of $-48.7\pm3.5$~µHz/s ($-2.6\times10^{-19}$/s) is measured at an optical power of $100$~nW.  

Our observation of negative and positive drift is in contrast to results
presented by Robinson et al. (Ref.~\onlinecite{Robinson2019}), where the drift
was always negative, regardless of the optical power. This discrepancy
may be due by the differences in the mirror coatings, as well as differences
in purity and internal stress of the mirror substrate material. Another
factor could be the length of time that the resonator was operated
at cryogenic temperature.

\section{Conclusion}

We have developed a relatively simple system for the stabilization
of the frequency of a laser on short- and medium time intervals, with
the goal of reaching a performance and a reliability comparable to
a hydrogen maser. The system is based on a $5$~cm long, vertically
oriented, silicon resonator operated at cryogenic temperature. The
system was characterized in detail. First, it is capable of continuous
operation. Except for realignment after initial cool-down, no significant
intervention is necessary when the resonator is operated either at
$1.5$~K or at $3.5$~K. Here we reported on over one year of data,
and characterization of the resonator's properties.

Second, the resonator vibration sensitivity was measured as $3\times10^{-10}$/$g$
in fractional terms. Together with vibrations produced by the cryostat
we expect this to be the limiting factor for the short-term frequency instability
of the resonator, which is equal to $1\times10^{-14}$ at $1$~s. 

At long integration times (1500~s) the instability is not
more than $2.5\times10^{-15}$, similar to the instability of the reference
hydrogen maser.

The long-term frequency drift rate was found to depend on the power
of the interrogating laser wave and on the duty cycle of the interrogation.
Our results, together with previous studies \cite{Hagemann2014, Wiens2016, Robinson2019}, indicate that in the limit of very low interrogation laser power and very low duty cycle the drift rate becomes extremely small.

By varying these parameters we could change the amplitude and sign
of the drift rate. The drift rate can be rather precisely modified
and this feature might in the future be used to produce a nearly drift-free
frequency reference.

We also characterized the temperature dependence of the resonator
frequency. We found two temperature values at which the frequency
has zero sensitivity with respect to temperature: $3.5$~K and $17.4$~K.
While the latter value is well known, the former is new. The much
smaller temperature derivative of the thermal frequency sensitivity
at $3.5$~K is highly advantageous, allowing to suppress temperature-induced
frequency instability below the Brownian noise for all integration
times. Moreover, the temperature value 3.5~K is sufficiently high
that it may be reached in cryostats not equipped with a Joule-Thomson
stage. This implies reduced complexity, purchase and maintenance costs.
Detailed studies are necessary to determine the origin of this promising
property. 

To improve the performance of our system, a cryostat that decouples vibrations 
of the cooler from the resonator is required. Such
cryostats are commercially available. We also note that the linewidth
of the resonator ($24.2$~kHz) could be lowered by replacing the
current mirrors with mirrors having lower loss; this could also improve
the performance of the system. In the future, a more precise characterization of
the instability could be done using as reference a high-performance
atomic standard (Cs clock or optical atomic clock).

\section{Funding}

E.W. acknowledges a fellowship from the Professor-W.-Behmenburg-Schenkung. This work was performed in the framework of project SCHI 431/21-1 of the Deutsche Forschungsgemeinschaft.

\begin{acknowledgments}
	We thank M. G. Hansen for his help with the operation of the frequency
	comb and A. Yu. Nevsky for stimulating discussions.
\end{acknowledgments}

%
\end{document}